\documentclass[
longbibliography,
showpacs,
letterpaper,
aps,
prc,
reprint,
superscriptaddress
]{revtex4-1}

\newcommand{\fig}[1]{Fig.~\ref{#1}}
\newcommand{\figs}[1]{Figs.~\ref{#1}} 

\newcommand{\tab}[1]{Table~\ref{#1}}
\newcommand{\tabs}[1]{Tables~\ref{#1}}

\newcommand{\rf}[1]{Ref.~\cite{#1}}

\newcommand{\sect}[1]{Sec.~\ref{#1}}

\usepackage{graphicx}
\usepackage{xspace}
\usepackage{ifthen}
\usepackage{array}
\usepackage{verbatim}
\usepackage{dcolumn}

\usepackage{lineno}
\def\linenumberfont{\normalfont\scriptsize\sffamily}

\newcommand{\id}{inner detector\xspace}
\newcommand{\nt}{neutrino target\xspace}
\newcommand{\gc}{gamma catcher\xspace}
\newcommand{\buf}{buffer\xspace}
\newcommand{\iv}{inner veto\xspace}
\newcommand{\ov}{outer veto\xspace}

\newcommand{\idns}{inner detector}
\newcommand{\ntns}{neutrino target}
\newcommand{\gcns}{gamma catcher}
\newcommand{\bufns}{buffer}
\newcommand{\ivns}{inner veto}
\newcommand{\ovns}{outer veto}

\renewcommand{\ol}{\begin{enumerate}}
\newcommand{\lo}{\end{enumerate}}

\newcommand{\ul}{\begin{itemize}}
\newcommand{\lu}{\end{itemize}}

\newcommand{\is} [1]{${}^{#1}$}
\newcommand{\ism} [1]{{}^{#1}}

\newcommand{\mup}{$\mu^+$\xspace}
\newcommand{\mum}{$\mu^-$\xspace}

\newcommand{\betap}{$\beta^{+}$\xspace}
\newcommand{\betam}{$\beta^{-}$\xspace}
\newcommand{\betan}{$\beta$n\xspace}

\newcommand{\hn}{\mbox{H-n}\xspace}
\newcommand{\gdn}{\mbox{Gd-n}\xspace}

\newcommand{\ba}[1]{{\boldmath \bf #1}}

\newcommand{\stab}[2]{---}
\newcommand{\fast}[2]{}
\newcommand{\slow}[2]{$\bullet$}
\newcommand{\isok}[2]{\is{#1}#2}
\newcommand{\rare}[2]{\emph{\is{#1}#2}}

\newcommand{\nrn}[3]{$\mathrm{#1}(\mu^-,\nu\mathrm{#2})\mathrm{#3}$}

\newcommand{\trn}[3]{$\mathrm{#1}(\mu^-,$&$\nu\mathrm{#2})\mathrm{#3}$}

\newcommand{\btwelveresfig}{\begin{figure}

\begin{center}
\includegraphics[width=\columnwidth]{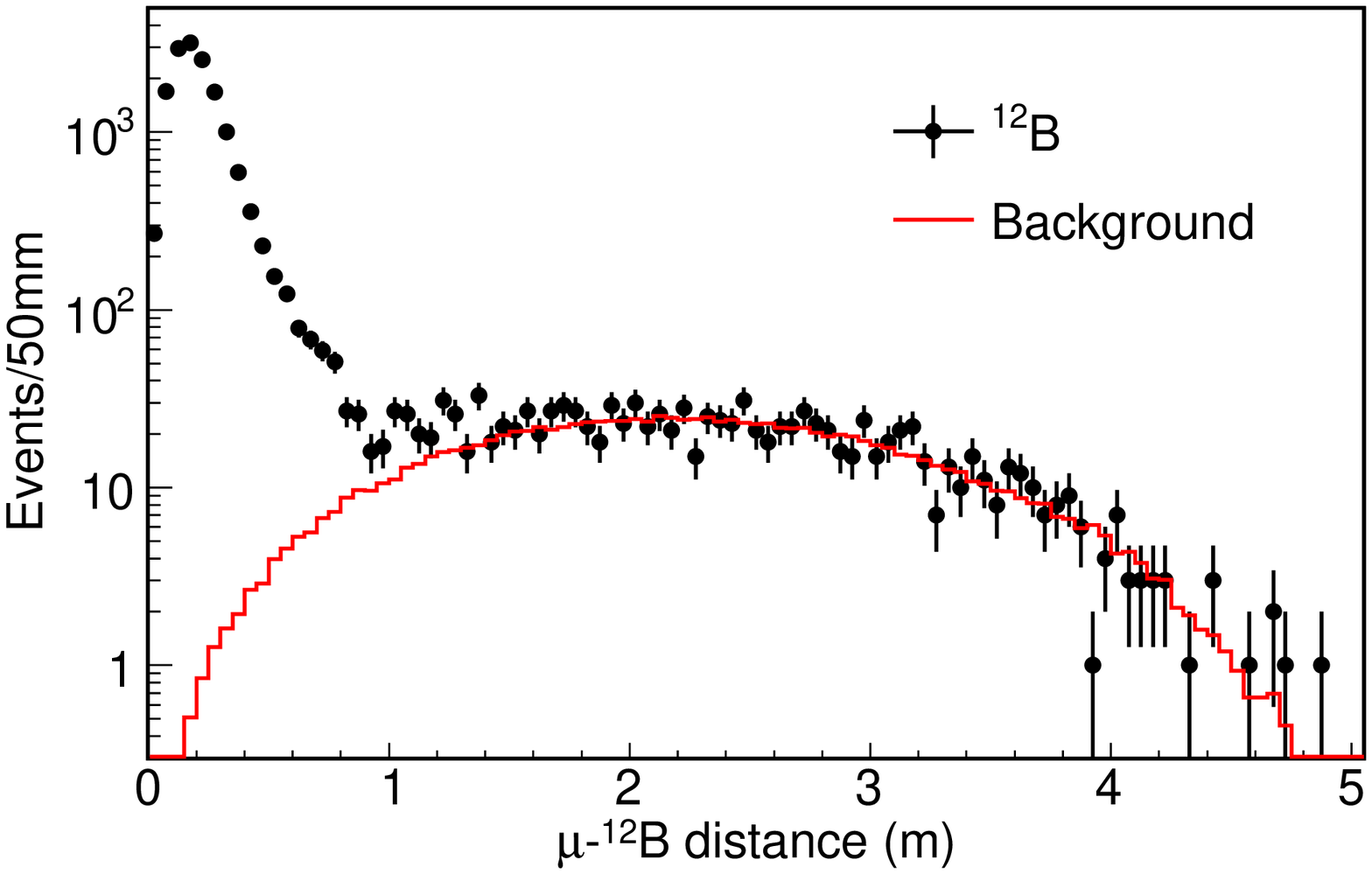}
\end{center}

\caption{\label{fig:b12res} Distance between selected \is{12}B events and
preceding stopped muon (black), as compared to a background sample (red).  The
efficiency for a distance cut is the fraction of background-subtracted \is{12}B
within the cut.}

\end{figure}}

\newcommand{\lininefig}{\begin{figure}
\includegraphics[width=\columnwidth]{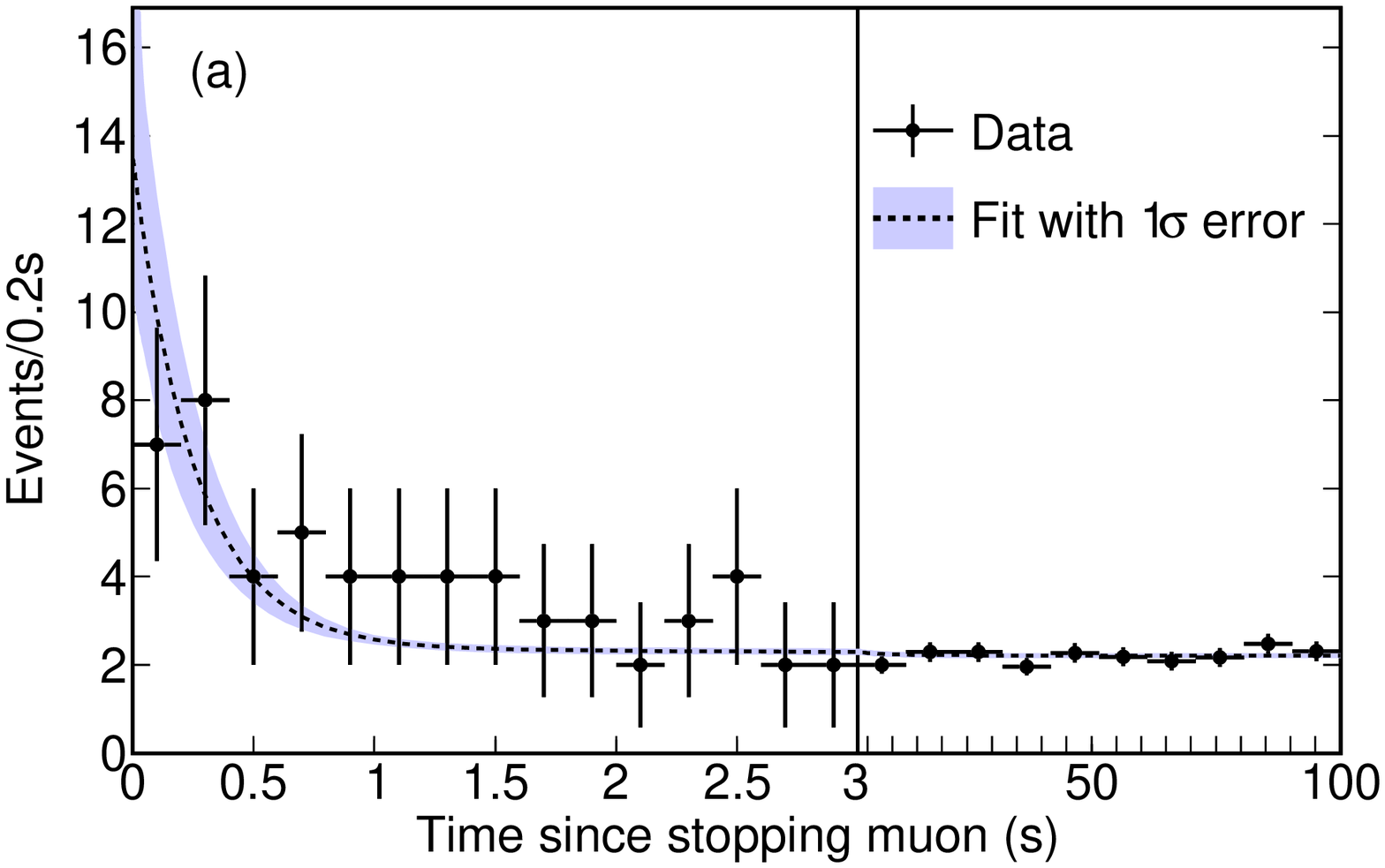}

\includegraphics[width=\columnwidth]{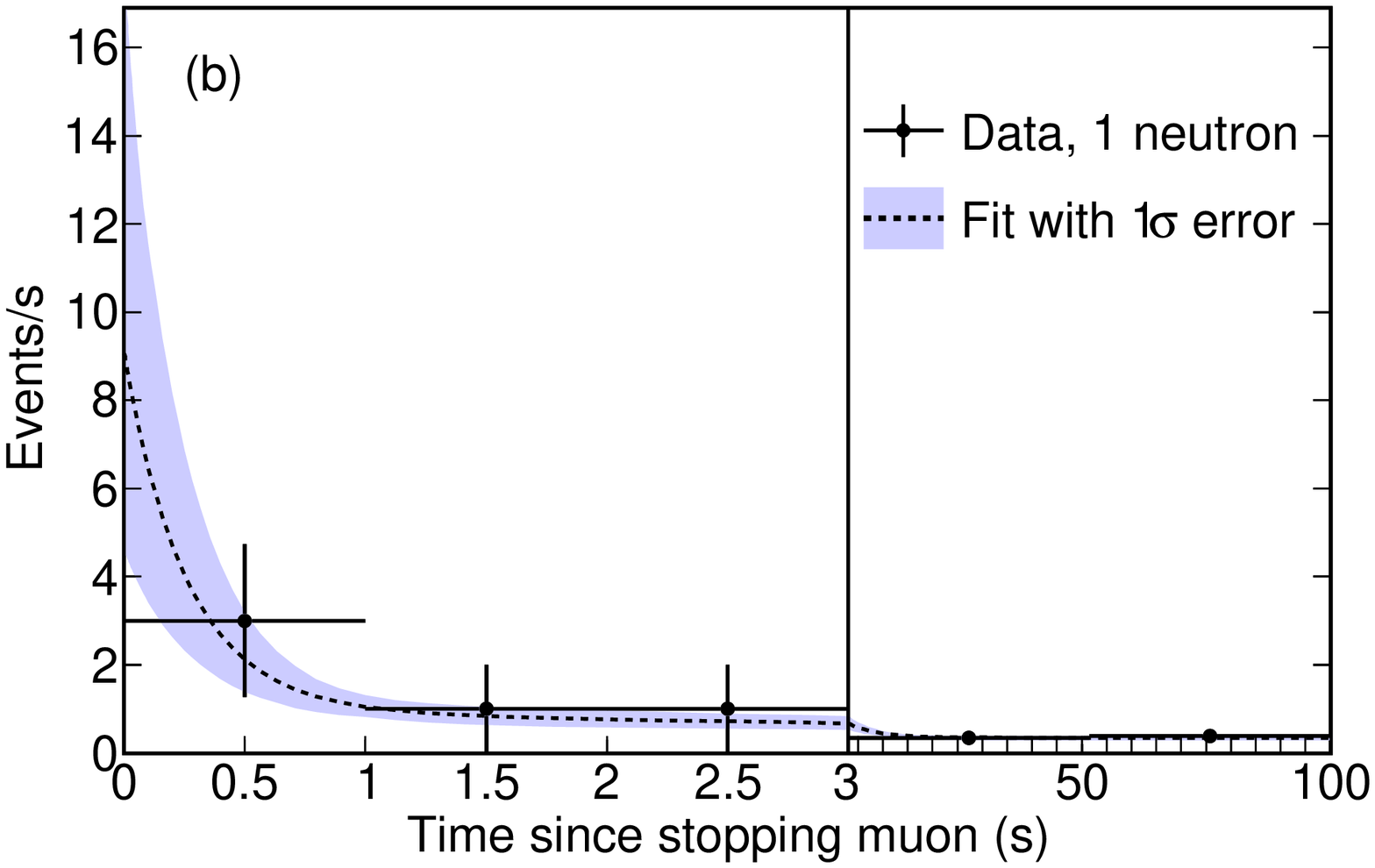}

\caption{\label{fig:li9} Observation of
\is{\mathrm{nat}}C$(\mu^-,\nu$X)\is{9}Li. (a) With no requirement for a
neutron following the muon. (b) Requiring exactly one neutron.
The fit results for \is{9}Li plus correlated (\is{16}C, \is{17}N) and
accidental ($\bar\nu_\mathrm e$) backgrounds  are shown. Note the change of
horizontal scale at 3\,s.}

\end{figure}}

\newcommand{\lininecontfig}{\begin{figure}

\begin{center}
\includegraphics[width=\columnwidth]{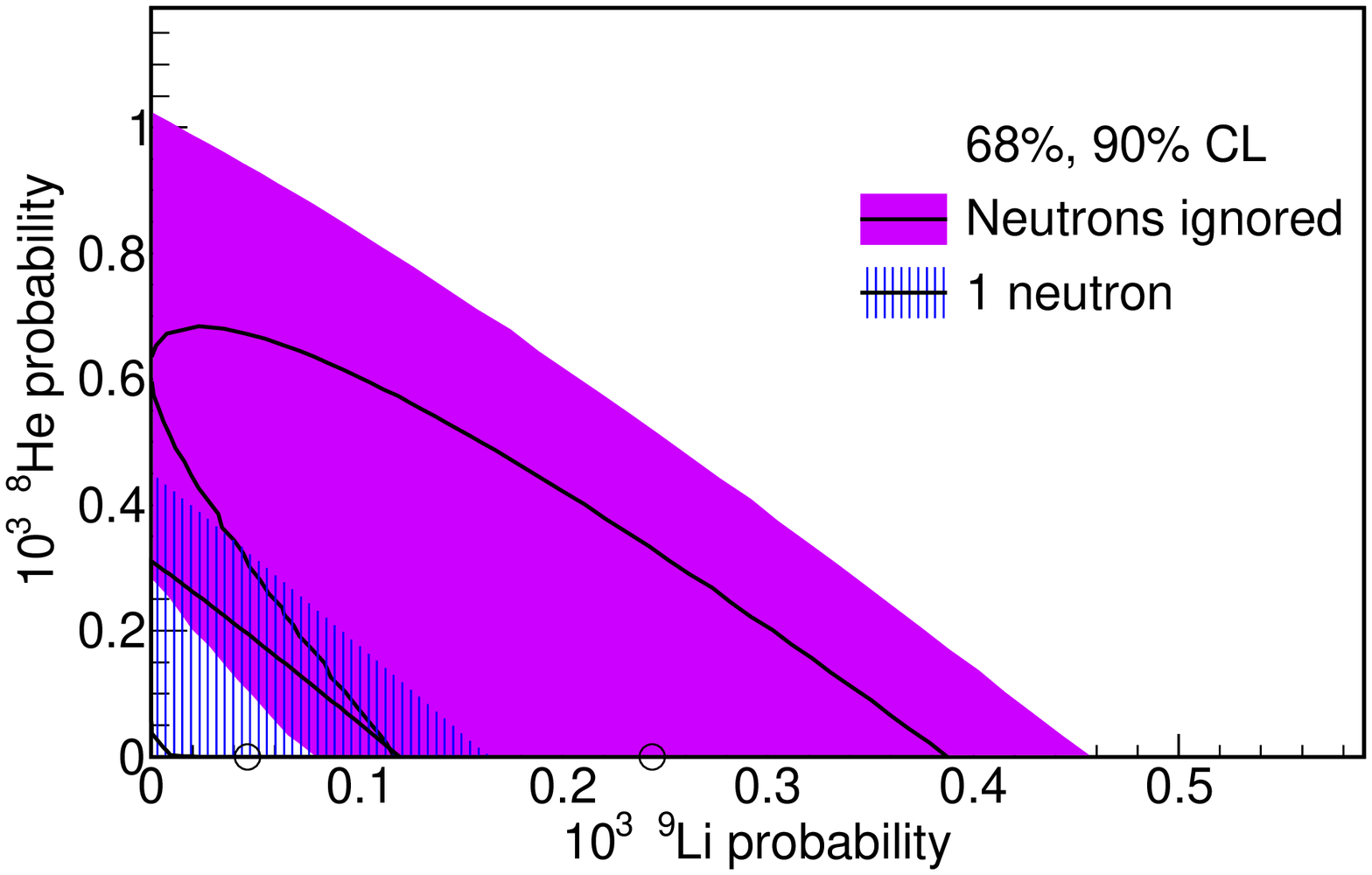}
\end{center}

\caption{\label{fig:li9cont} 90\% and 68\% contours for \is{9}Li and
\is{8}He with no neutron requirement (violet) and requiring a neutron
(blue), with probabilities given per \is{\mathrm{nat}}C capture.}

\end{figure}}

\newcommand{\linineenergyfig}{\begin{figure}

\begin{center}
\includegraphics[width=\columnwidth]{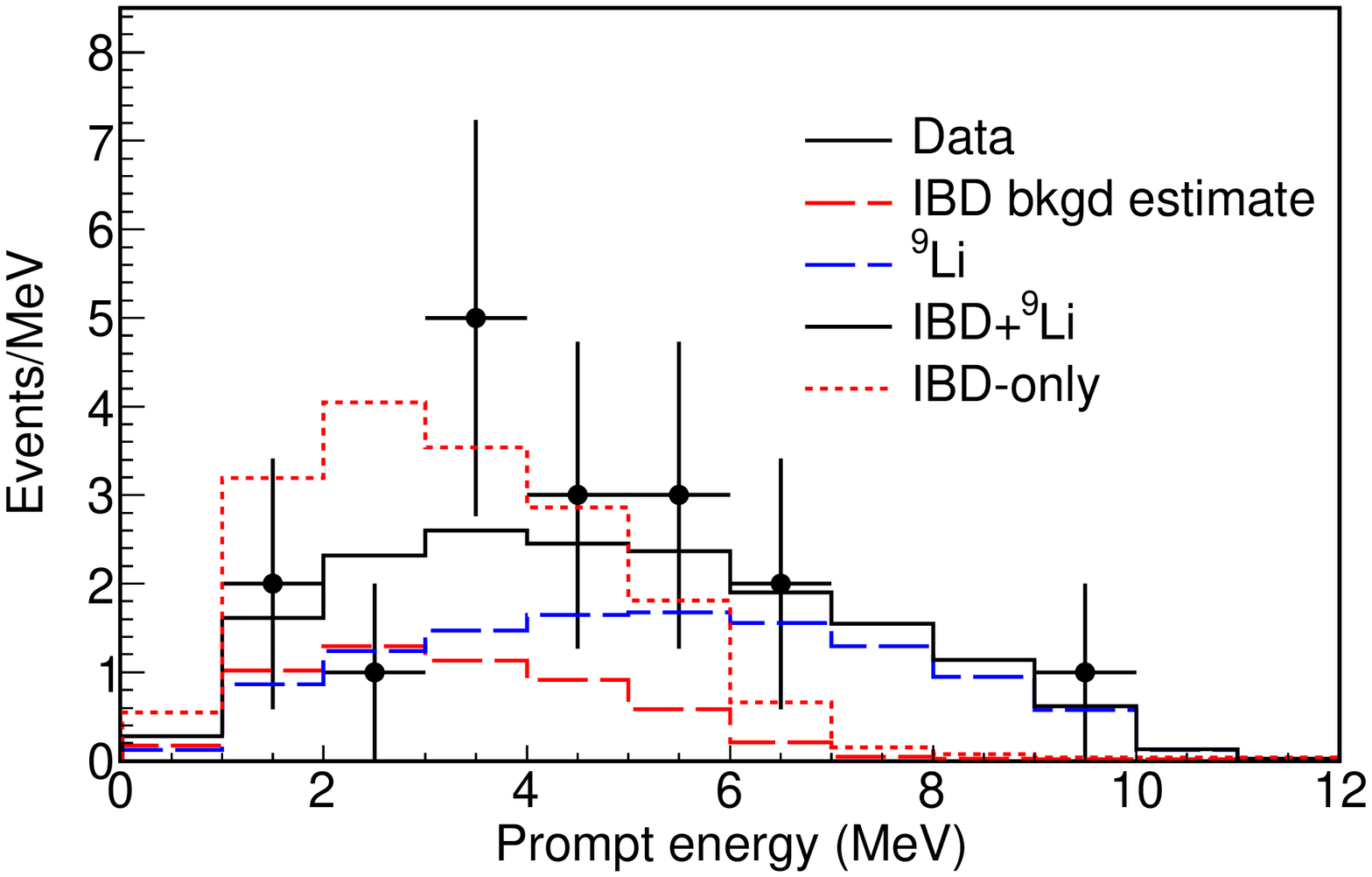}
\end{center}

\caption{\label{fig:li9energy} Prompt energy for \is{9}Li candidates with
$\Delta t < 0.5\,$s.  An IBD background estimate from offtime samples is shown
in dashed red.  MC for \is{9}Li is shown in dashed blue, normalized so that the
sum (solid black) matches the data. In dashed red is shown  the IBD-only
spectrum normalized to the data. The highest energy event has 9.47\,MeV.}

\end{figure}}

\newcommand{\bthirteenlielevenfig}{\begin{figure}
\begin{center}
\includegraphics[width=\columnwidth]{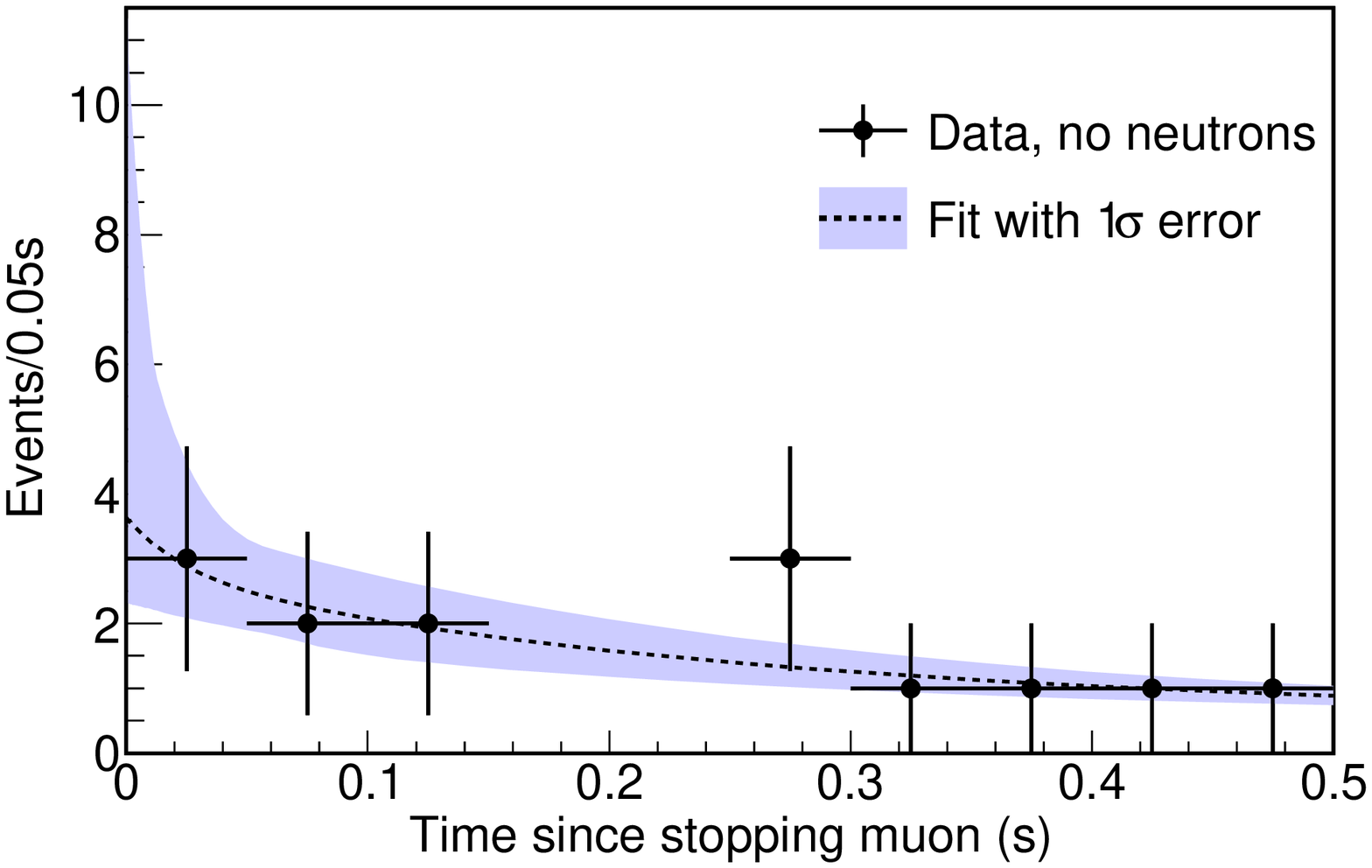}
\end{center}

\caption{\label{fig:b13li11} Search for \is{11}Li and the $\beta$n mode of
\is{13}B.  The dashed line shows the best fit, which finds no significant
amount of \is{11}Li or \is{13}B.}

\end{figure}}

\newcommand{\btwelvefromcthirteenfig}{\begin{figure}
\begin{center}
\includegraphics[width=\columnwidth]{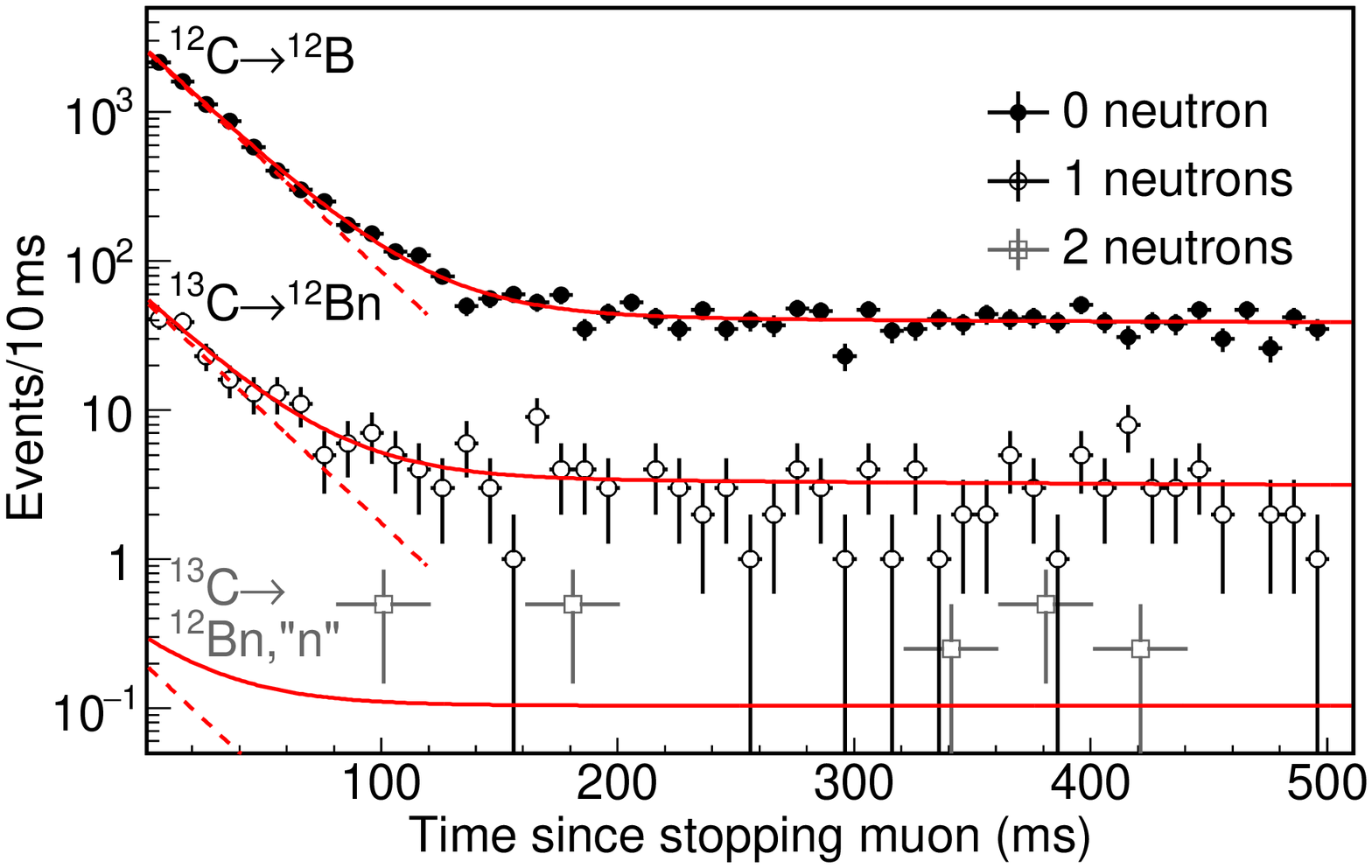}
\end{center}

\caption{\label{fig:b12fromc13} \is{12}B production with and without a neutron,
interpreted as \is{13}C$(\mu^-,\nu)$\is{12}B (filled circles),
\is{13}C$(\mu^-,\nu n)$\is{12}B (open circles).  Neutron efficiency and
accidental neutron-like events are accounted for, including those occurring
along with a real neutron (open squares; ``n'' denotes an accidental
neutron-like event). The solid red curves show the fit results, while the
dashed red lines give the \is{12}B components thereof.}

\end{figure}}

\newcommand{\lieightfig}{\begin{figure}
\begin{center}
\includegraphics[width=\columnwidth]{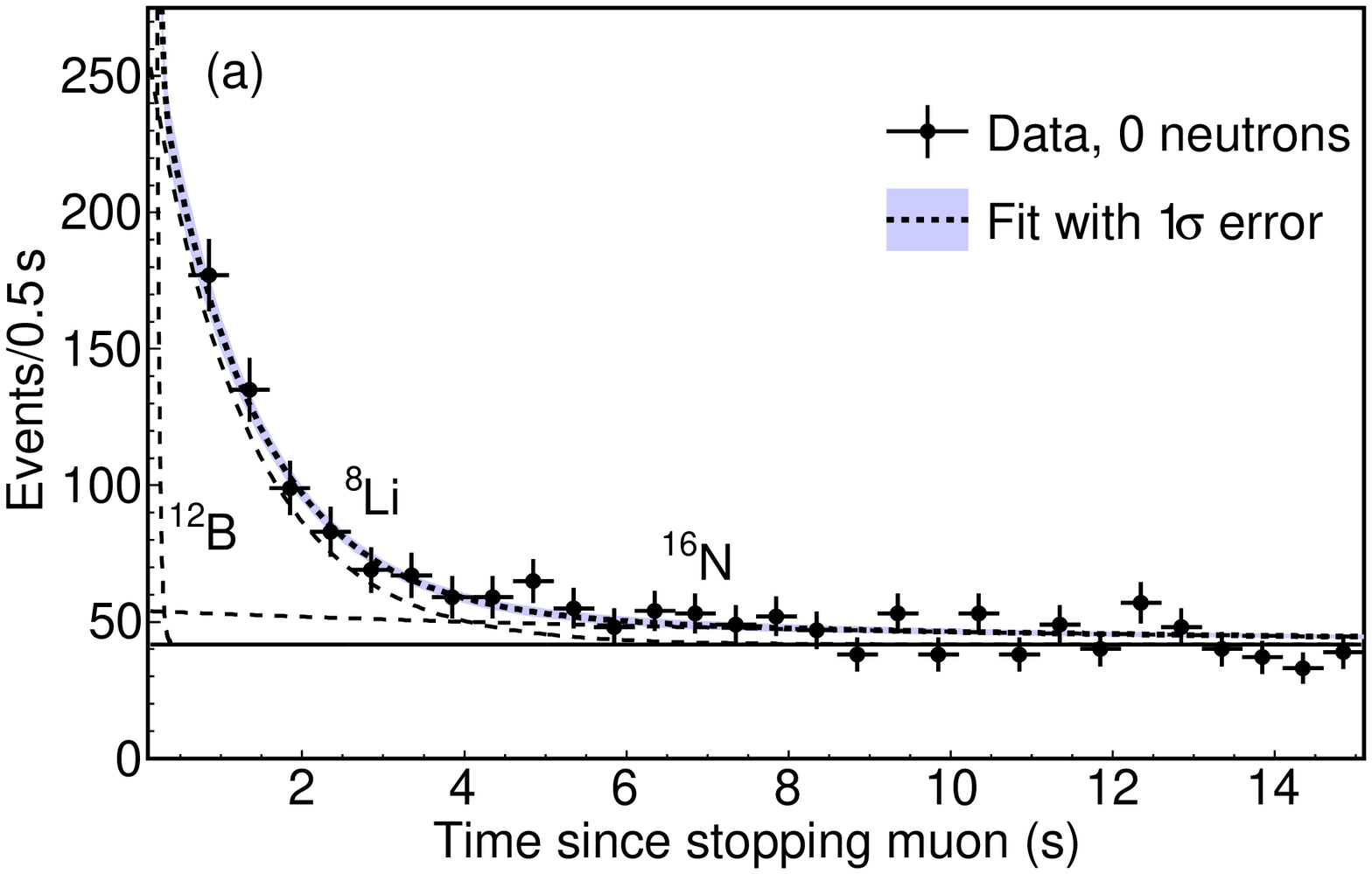}

\includegraphics[width=\columnwidth]{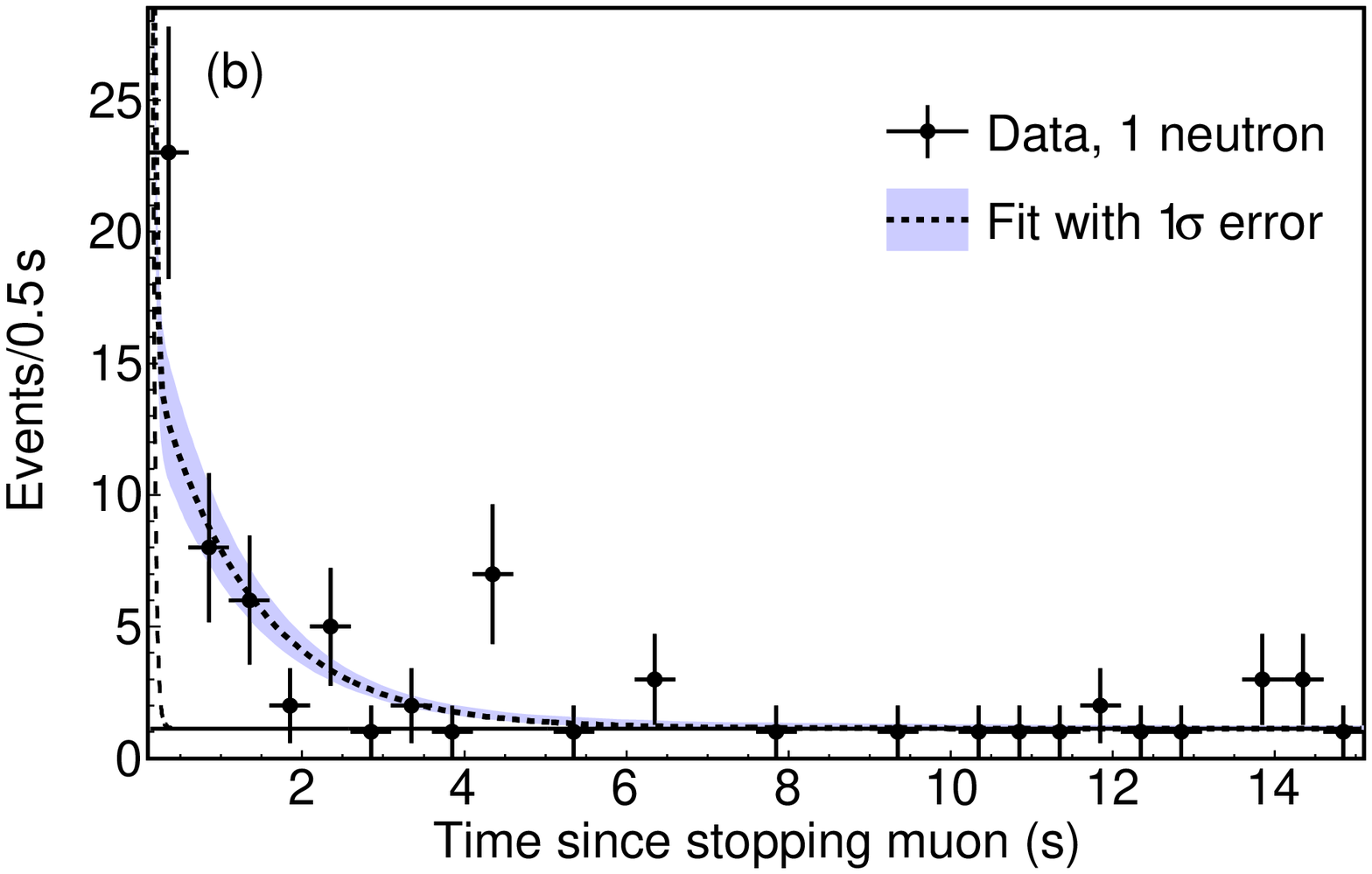}

\includegraphics[width=\columnwidth]{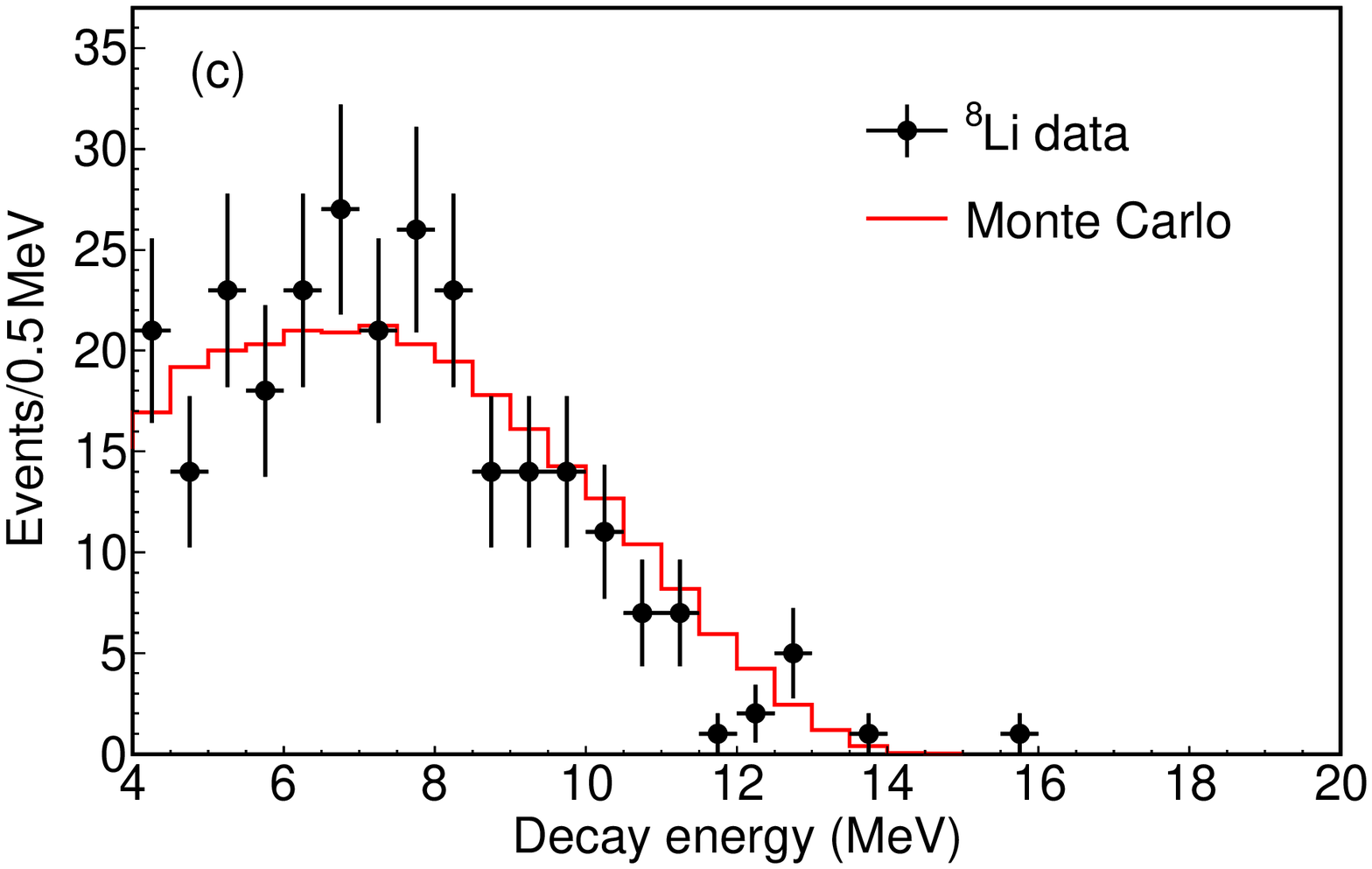}
\end{center}

\caption{\label{fig:li8} \is{8}Li search. (a) Joint fit for \is{12}B,
\is{8}Li, \is{16}N and accidentals when no neutrons follow the muon. (b)
Same, with one neutron. (c) Energy of events in the zero-neutron case with
a tighter cut of \mbox{$\Delta r < 200$\,mm} when $0.3\,\mathrm{s} < \Delta t <
2\,$s (points).  The solid line shows Monte Carlo normalized to the data.}

\end{figure}}

\newcommand{\beelevenfig}{\begin{figure}
\begin{center}
\includegraphics[width=\columnwidth]{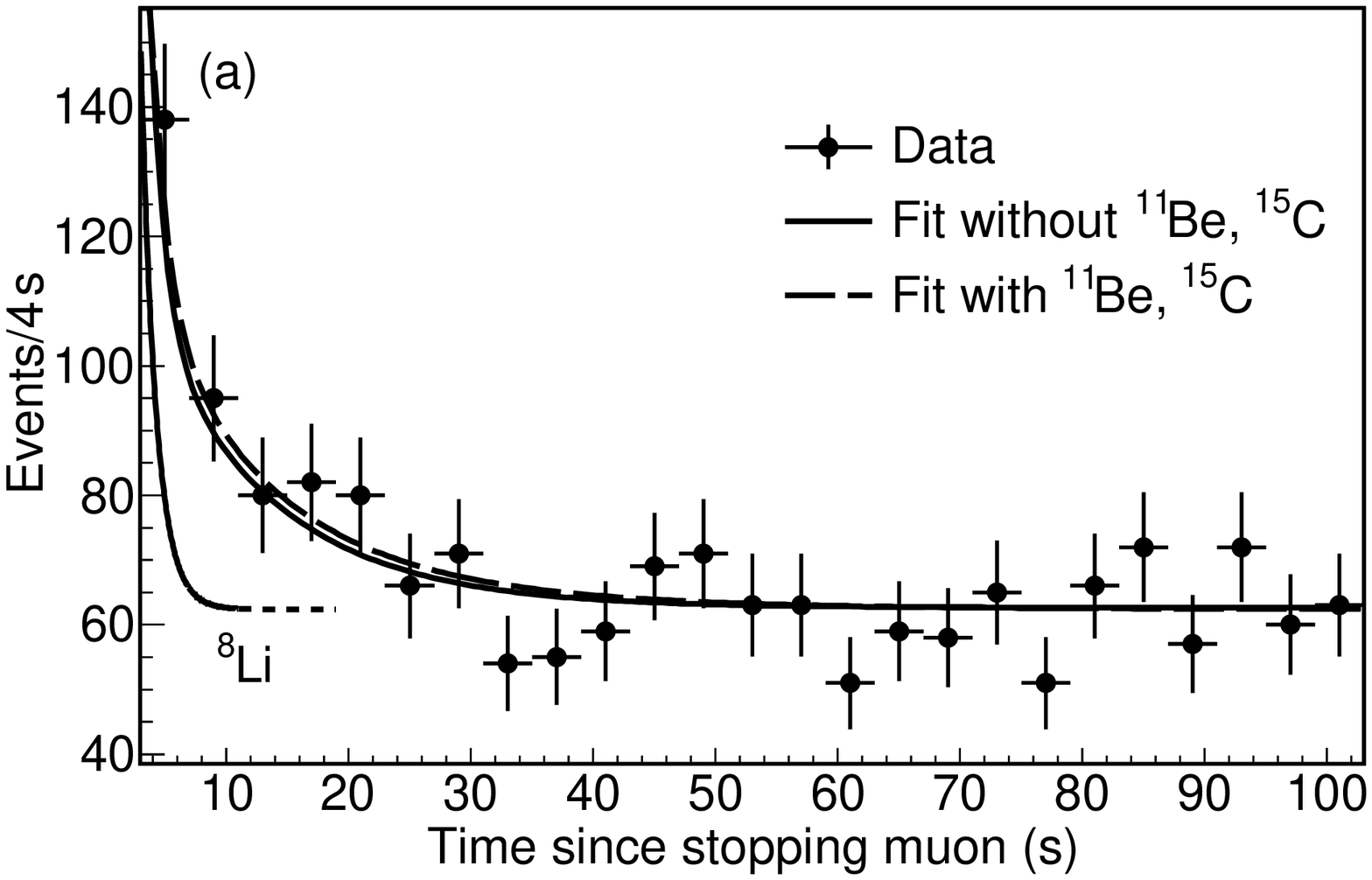}

\includegraphics[width=\columnwidth]{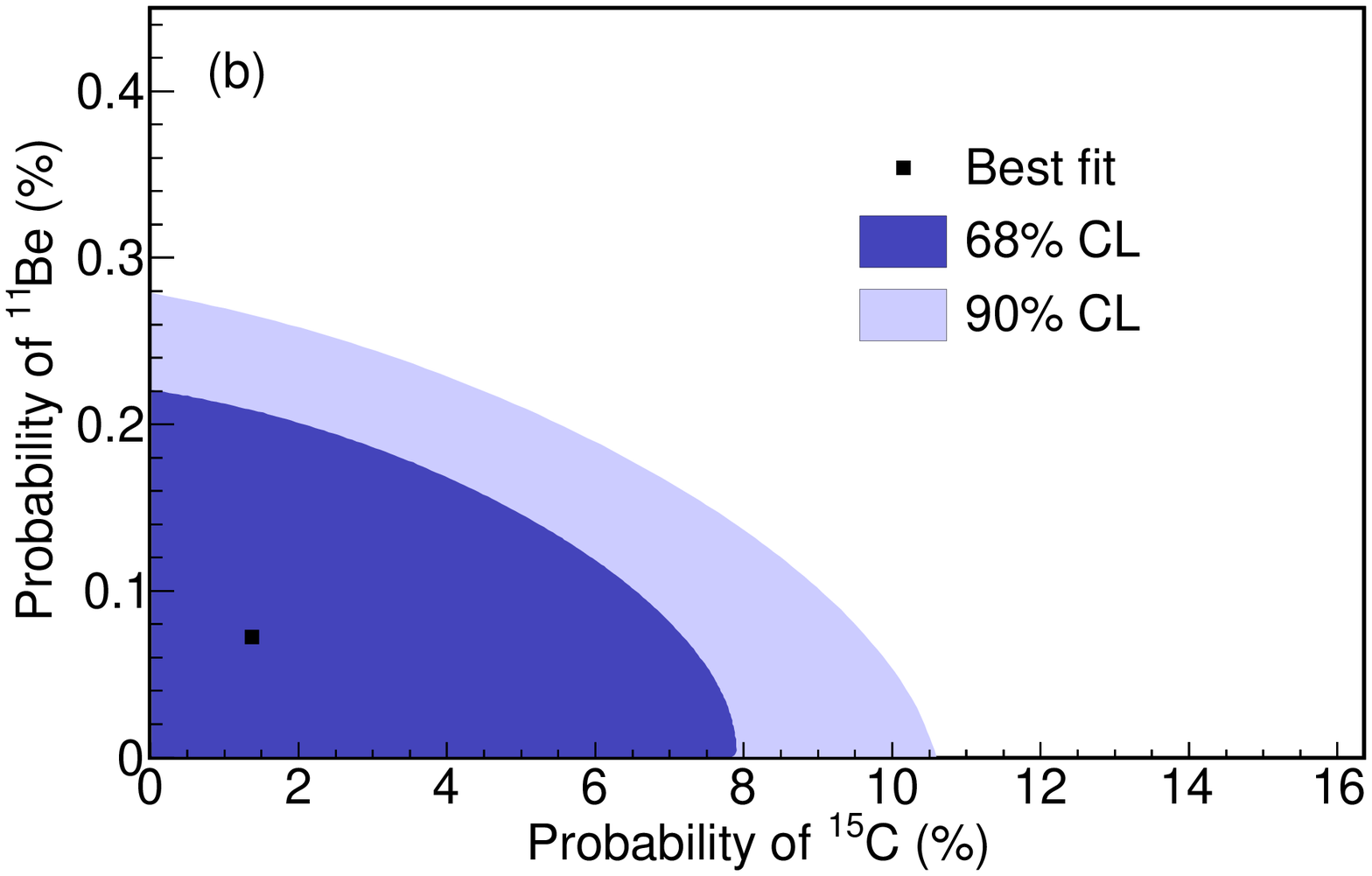}
\end{center}

\caption{\label{fig:be11} (a) Search for \is{11}Be and \is{15}C.  The solid
(dashed) line shows the best fit without (with) \is{11}Be or \is{15}C.  The
lower curve shows the contribution of \is{8}Li; all contribution from \is{12}B
is to the left of the plot area due to its short half-life. (b)
Statistics-only confidence regions for \is{15}C and \is{11}Be.  Blue (tan)
shows 68\% (90\%) CL.  }

\end{figure}}

\newcommand{\hesixfig}{\begin{figure}
\begin{center}
\includegraphics[width=\columnwidth]{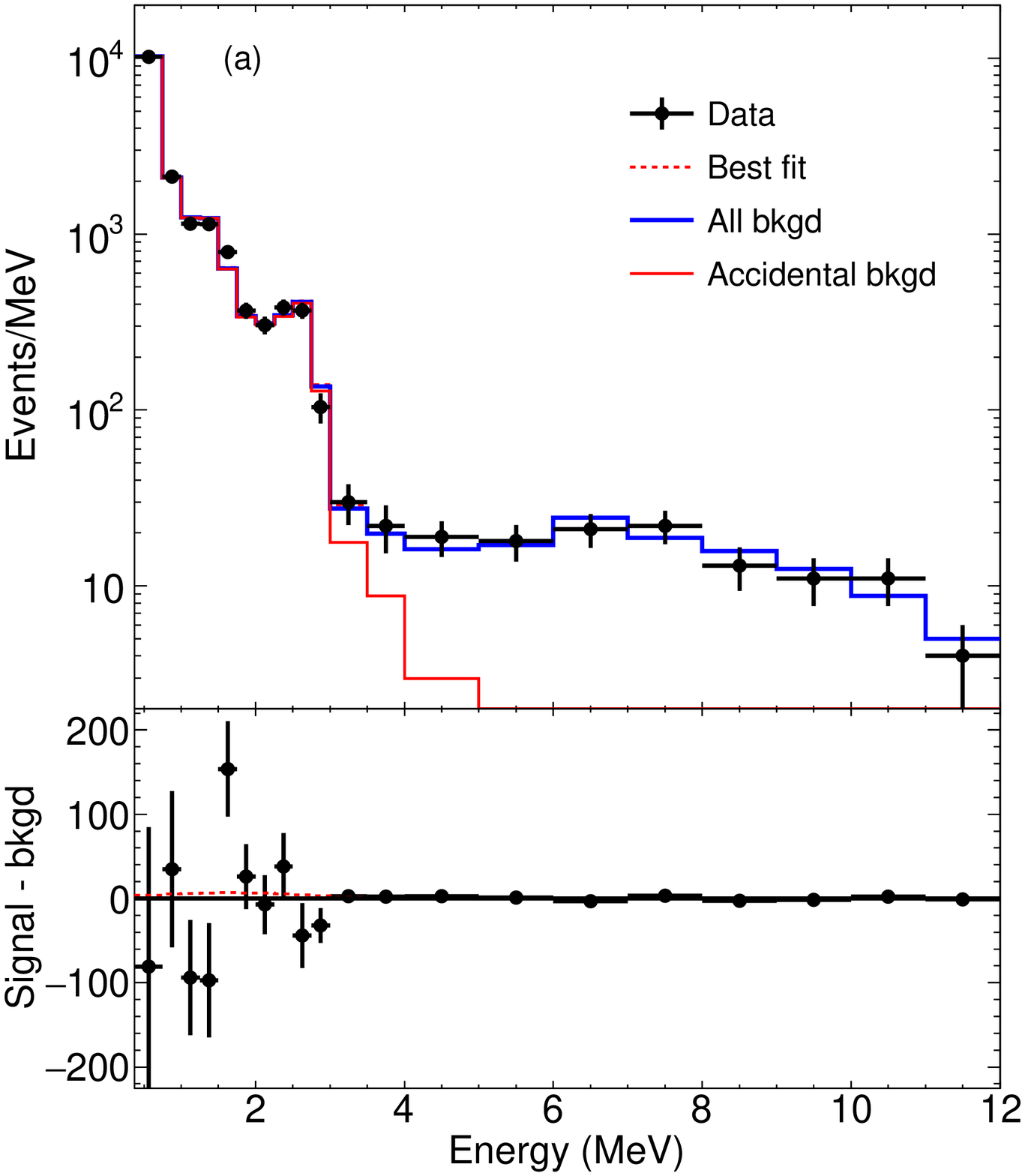}

\includegraphics[width=\columnwidth]{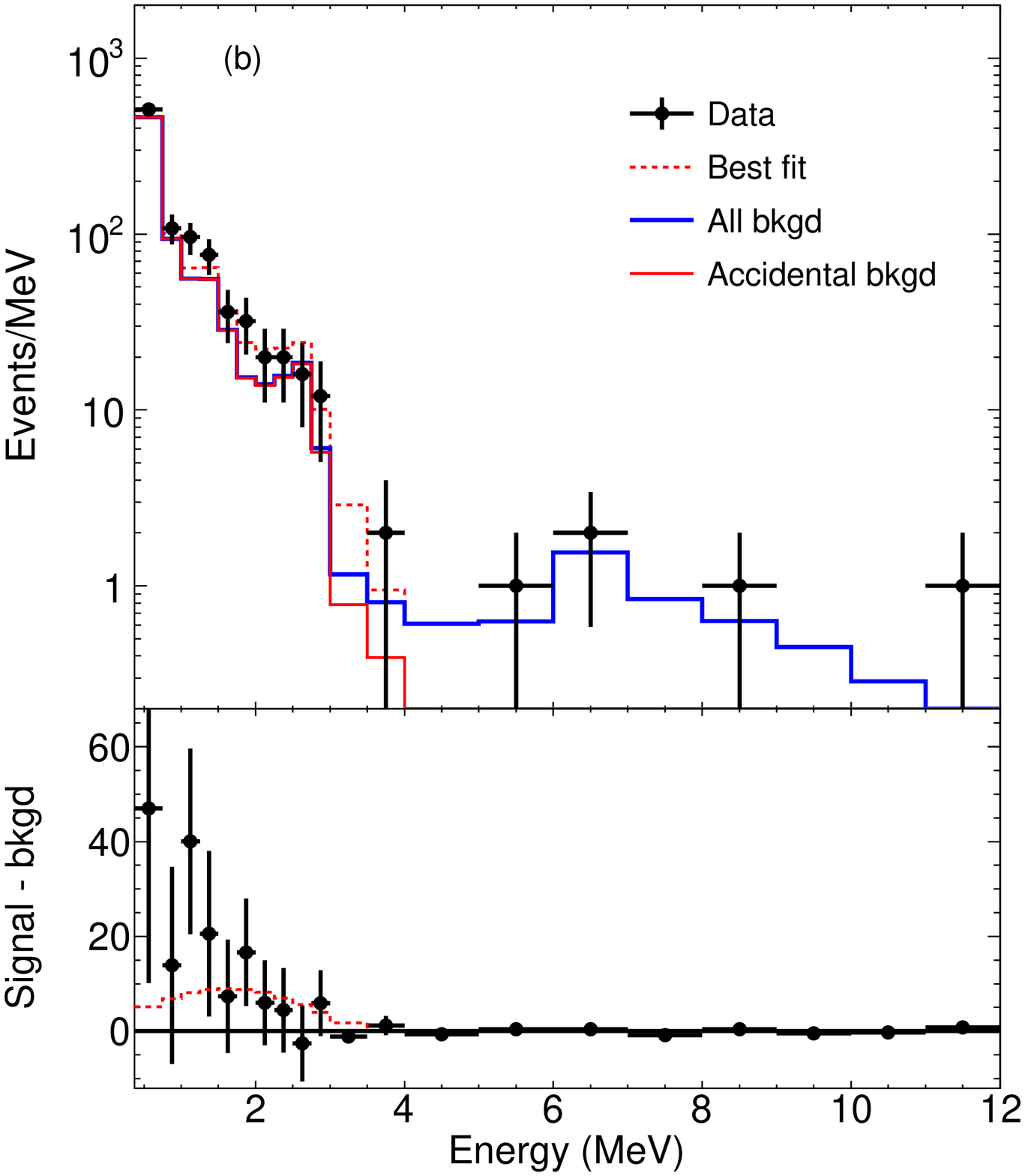}
\end{center}

\caption{\label{fig:he6} (a) Fit for \is{6}He with zero 
neutrons. (b) Same, for one neutron. This is a display binning that
combines the four \nt regions.
The accidental background is shown in solid red.  Stacked on this is the
correlated background (\is{8}Li and \is{16}N) in solid blue. The best fit
including \is{6}He is shown in dotted red.}

\end{figure}}

\newcommand{\hensixfig}{\begin{figure}
\begin{center}
\includegraphics[width=\columnwidth]{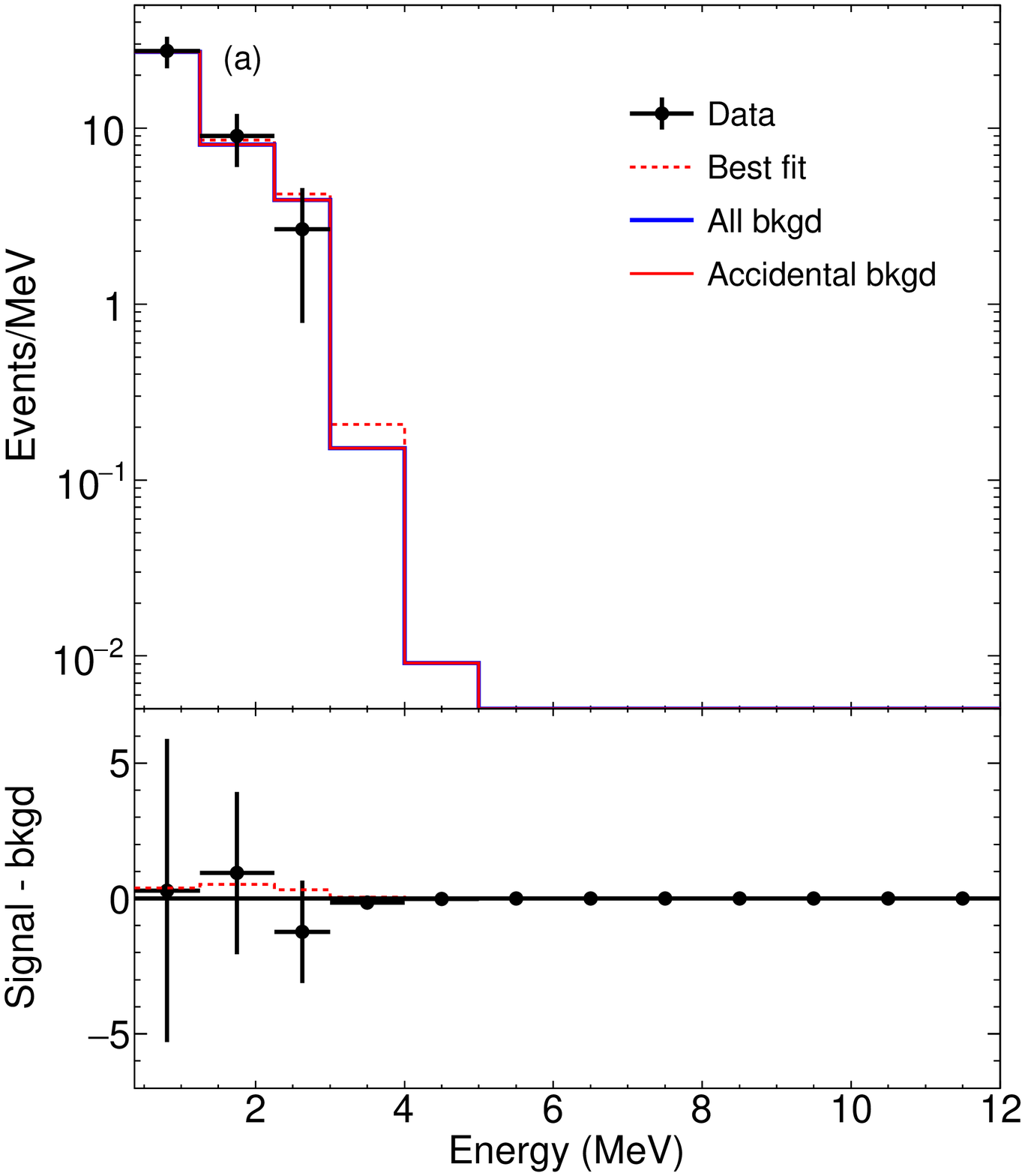}

\includegraphics[width=\columnwidth]{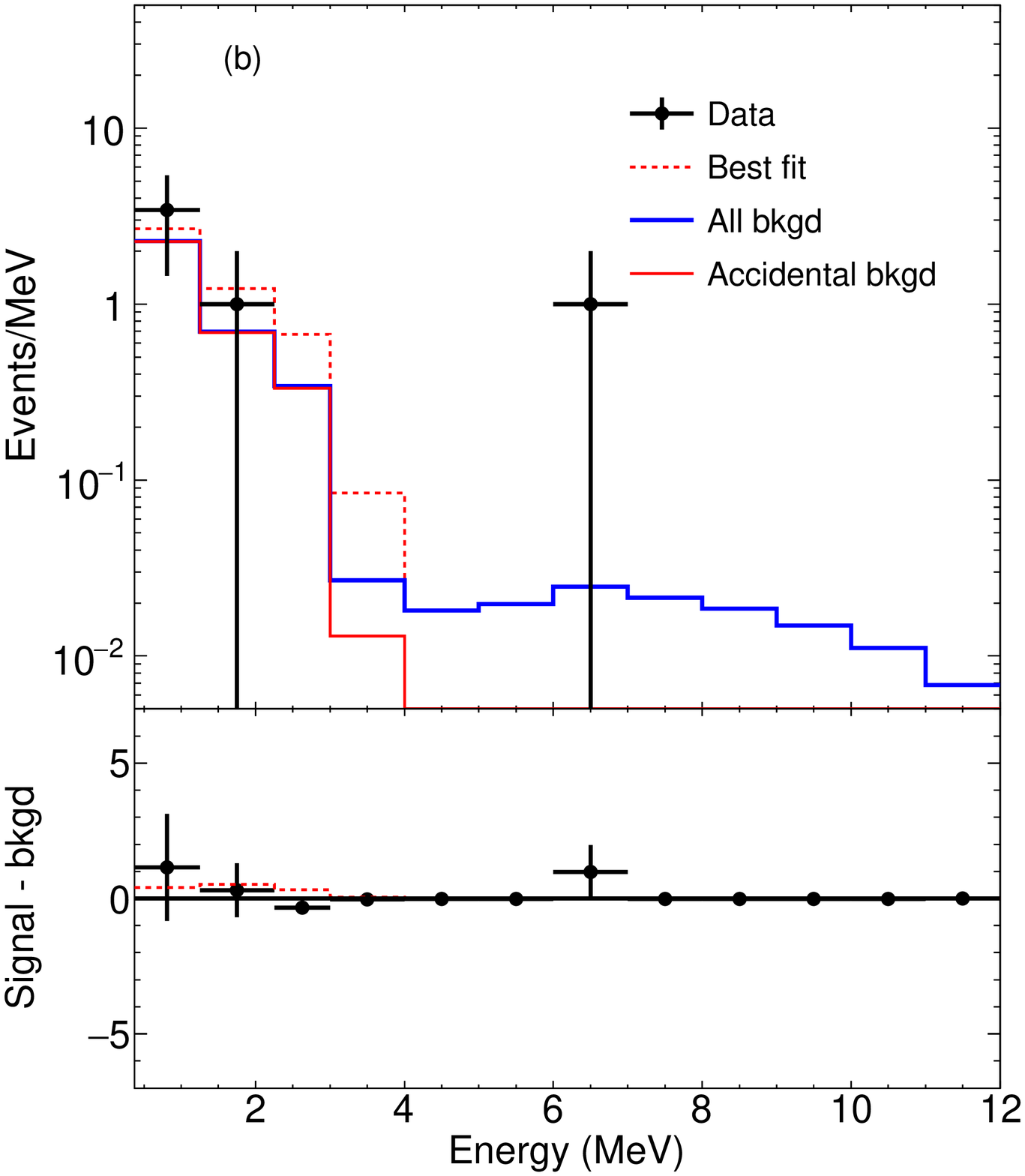}
\end{center}

\caption{\label{fig:he6nn} (a) Fit for \is{6}He with two 
neutrons. (b) Same, for three neutrons. The conventions are
the same as in \fig{fig:he6}. }

\end{figure}}

\newcommand{\lininedposfig}[1][tbp]{\begin{figure}
\begin{center}
\includegraphics[width=\columnwidth]{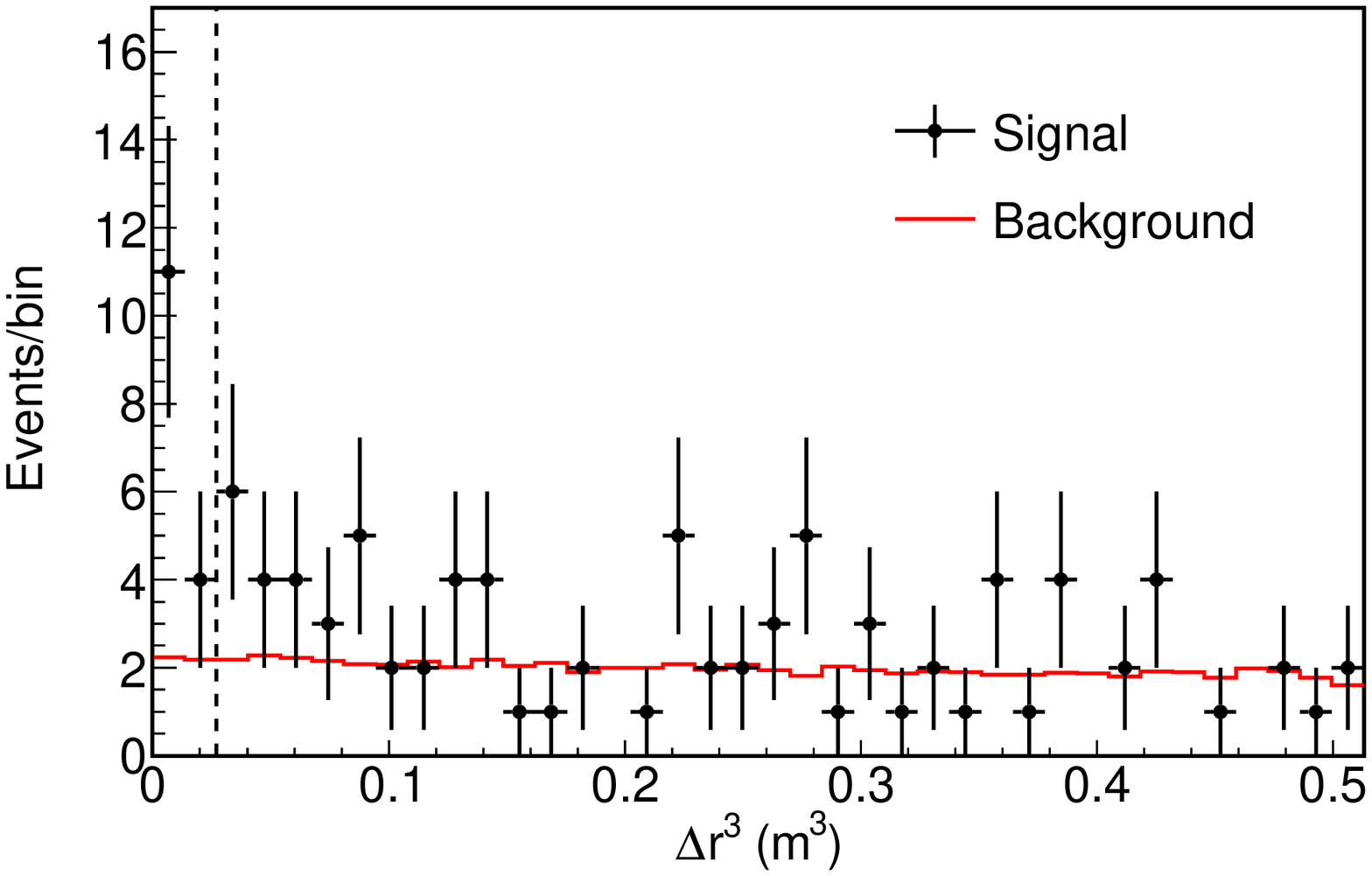}
\end{center}

\caption{\label{fig:li9dpos} The distance-cubed distribution of \betan events
less than 0.4\,s after a stopping muon with all other cuts applied. The dashed
line is the $\Delta r$ cut.}

\end{figure}}

\newcommand{\bfourteenfig}[1][tbp]{\begin{figure}[#1]
\begin{center}
\includegraphics[width=\columnwidth]{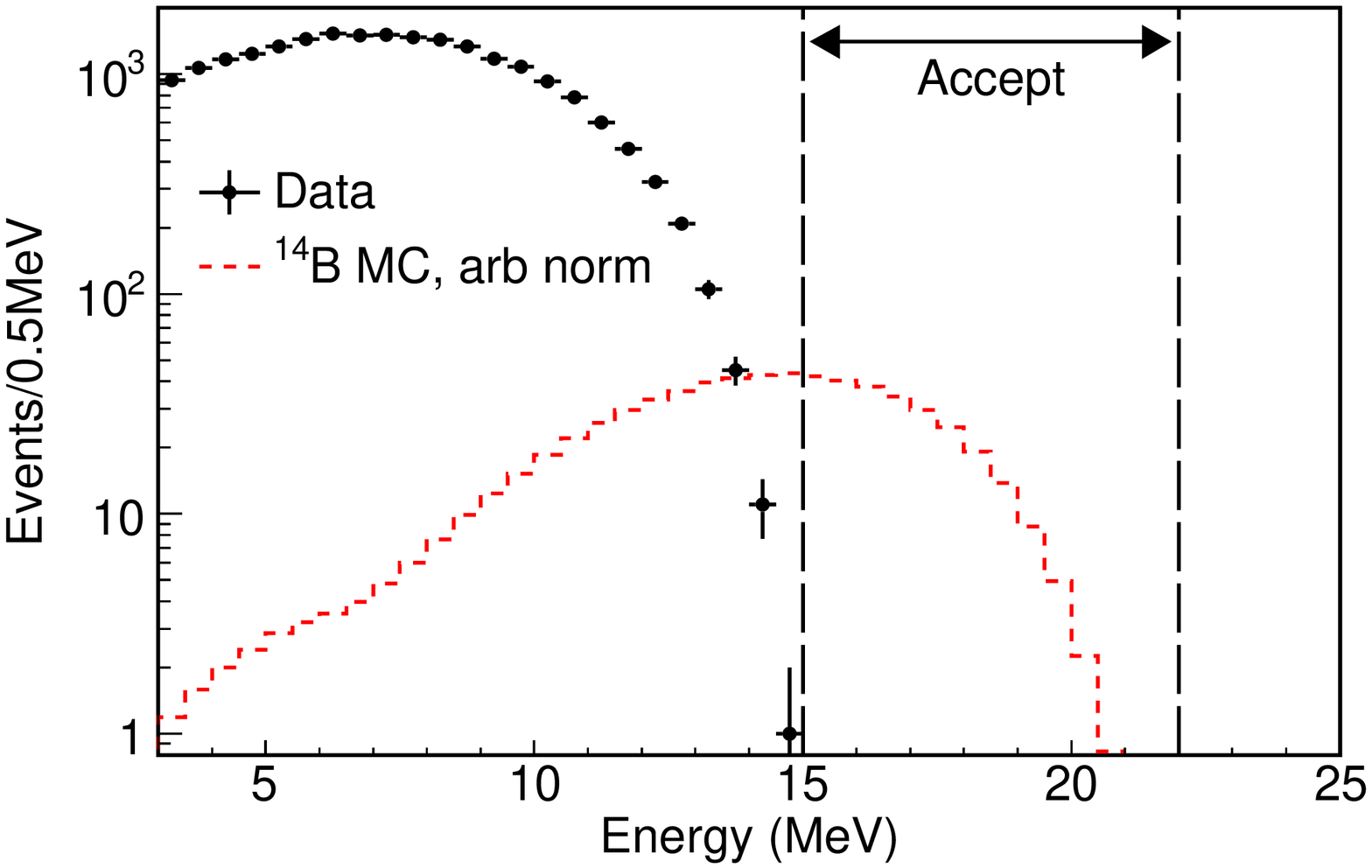}
\end{center}

\caption{\label{fig:b14} Energy cut for \is{14}B search, showing no events in
the signal region. The data below the cut is primarily \is{12}B.}

\end{figure}}

\newcommand{\btwelvegammafitfig}{\begin{figure*}

\begin{center}
\includegraphics[width=\columnwidth]{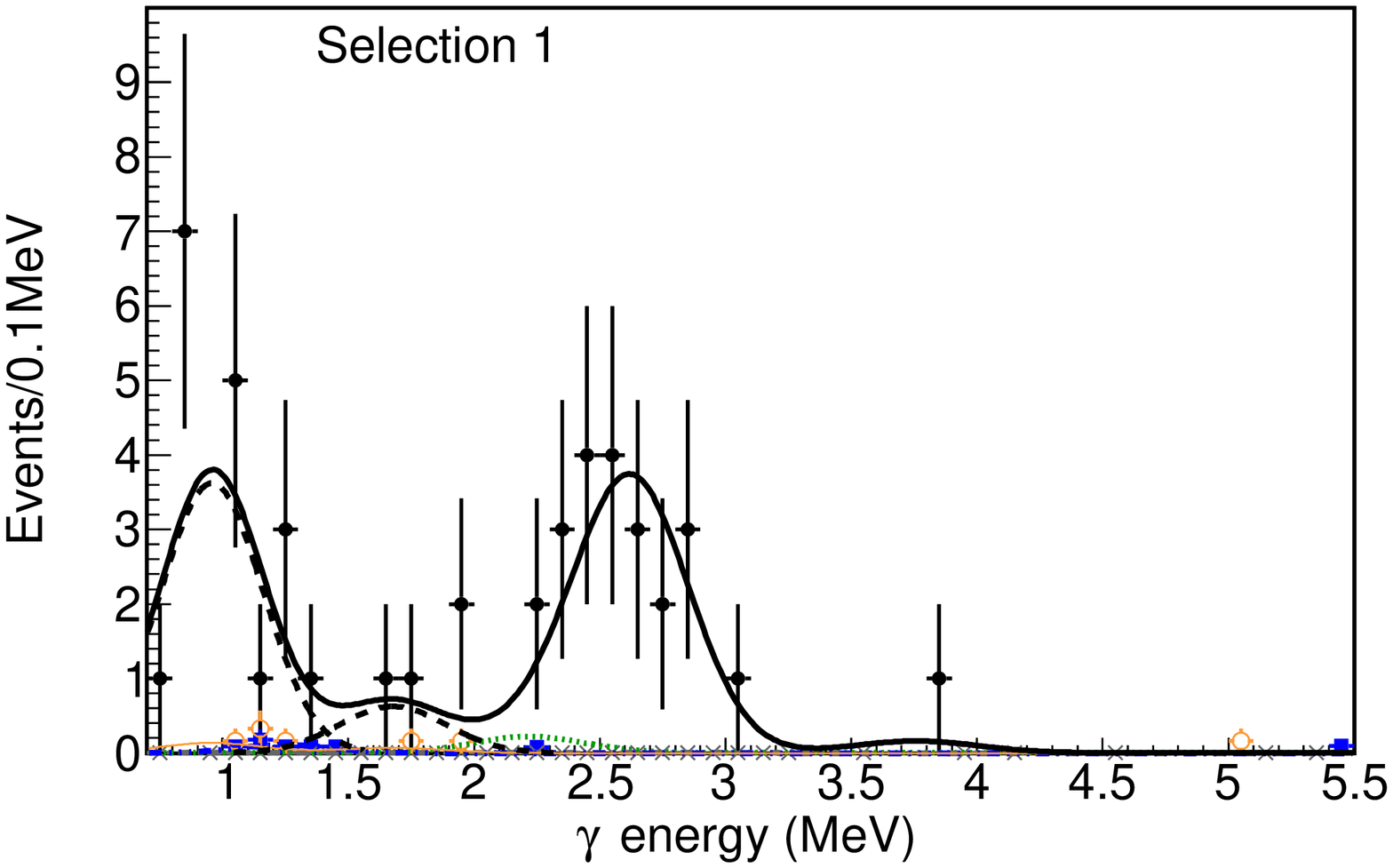}
\includegraphics[width=\columnwidth]{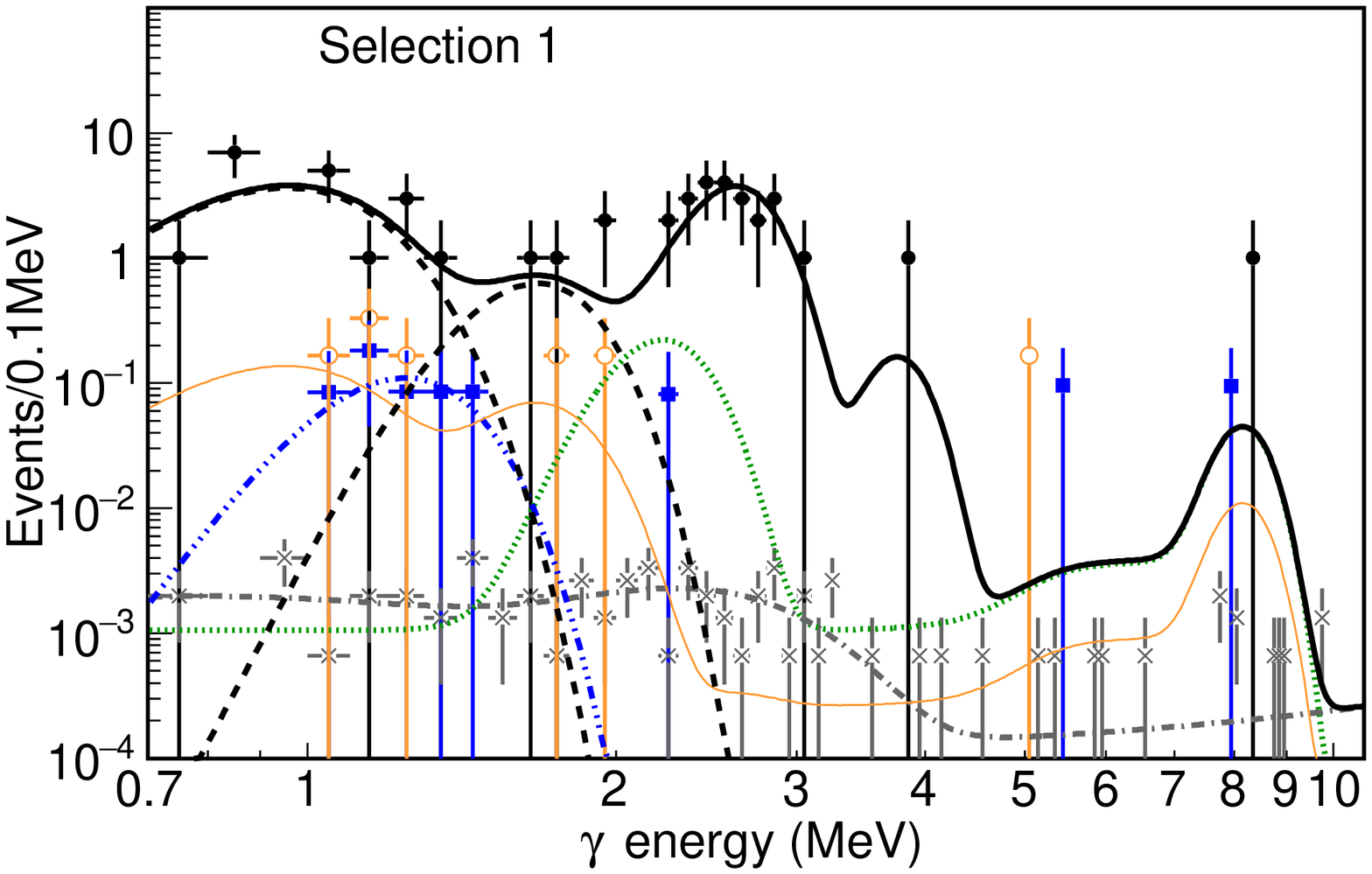}

\includegraphics[width=\columnwidth]{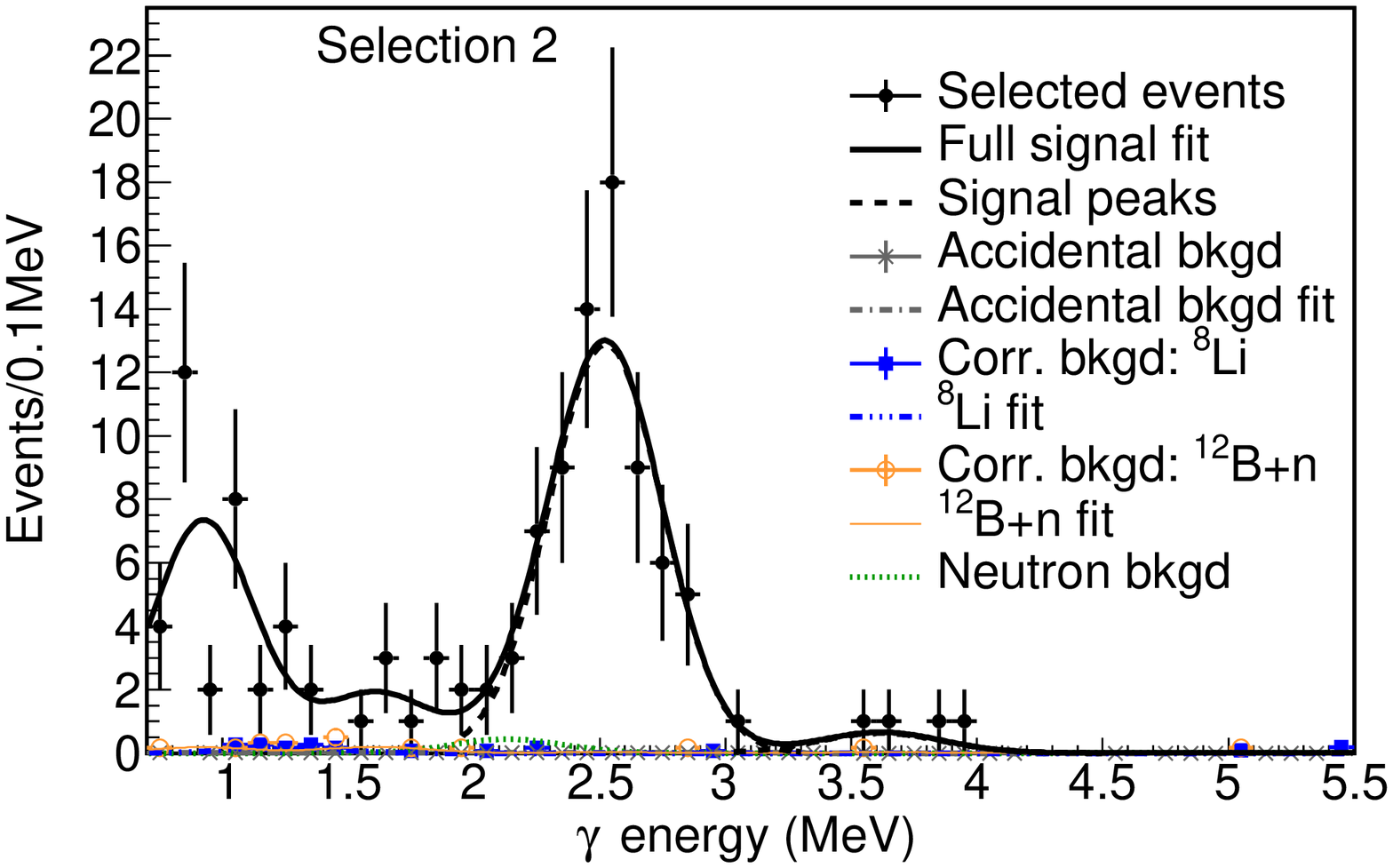}
\includegraphics[width=\columnwidth]{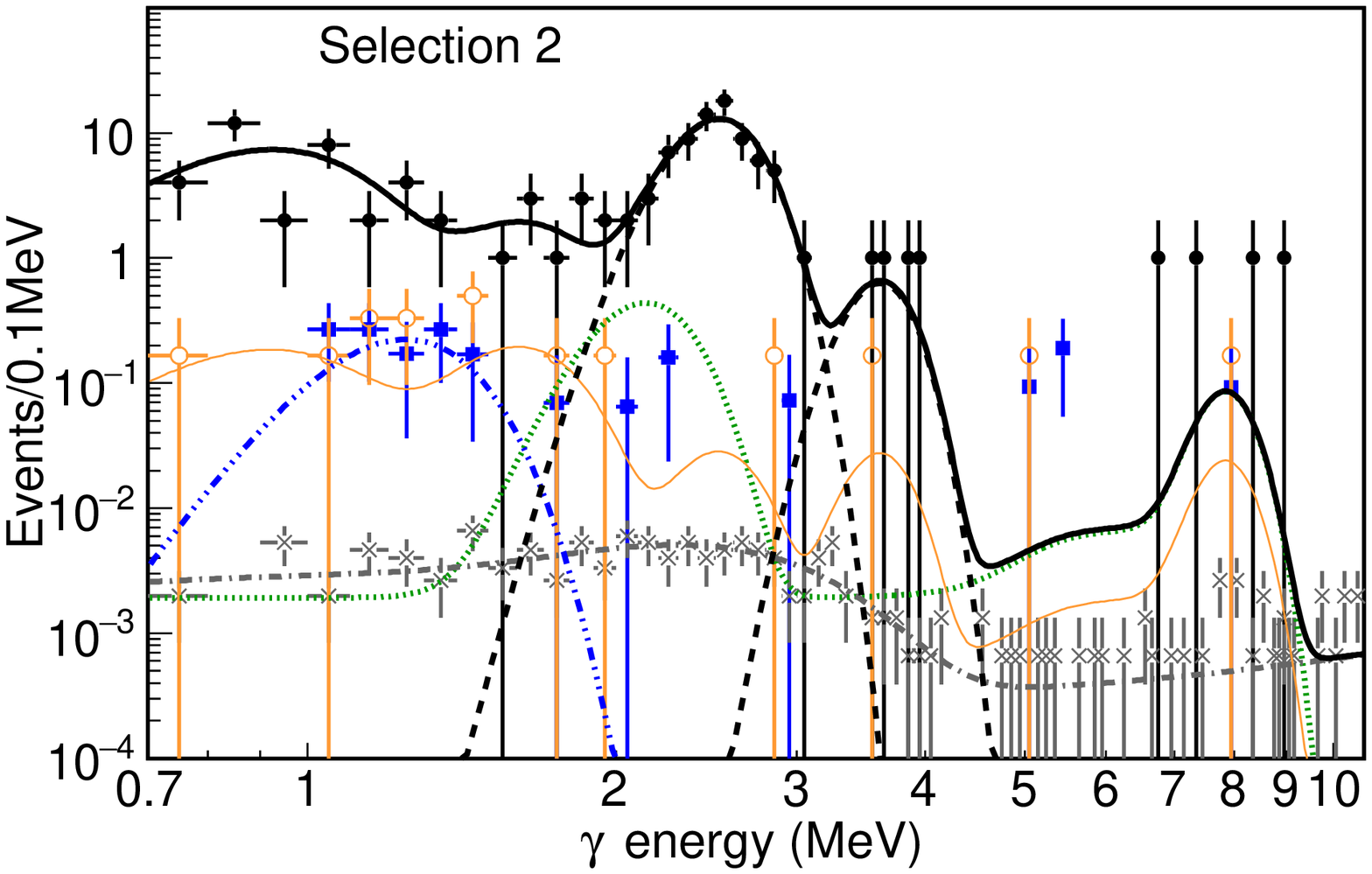}

\includegraphics[width=\columnwidth]{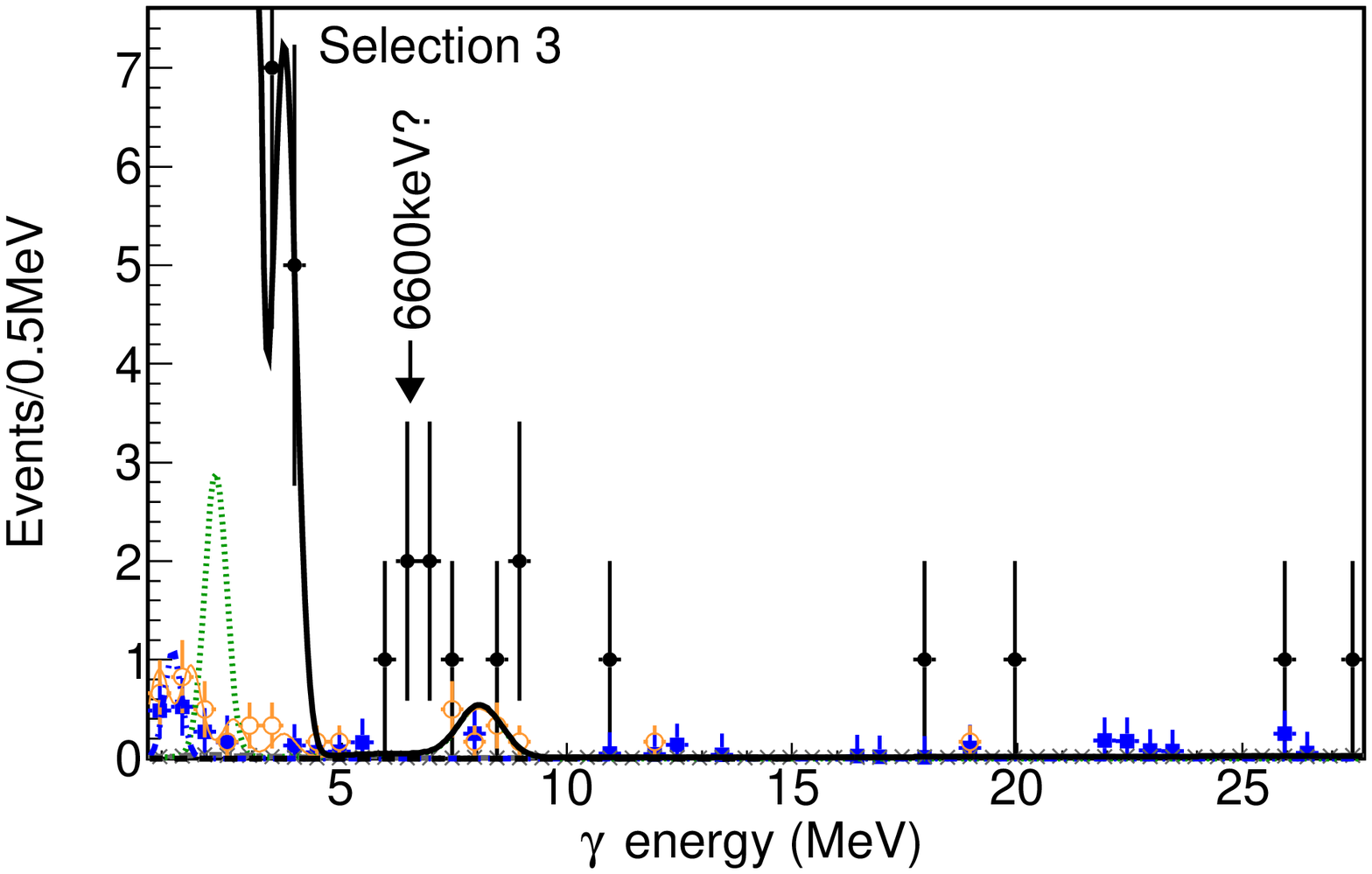}
\includegraphics[width=\columnwidth]{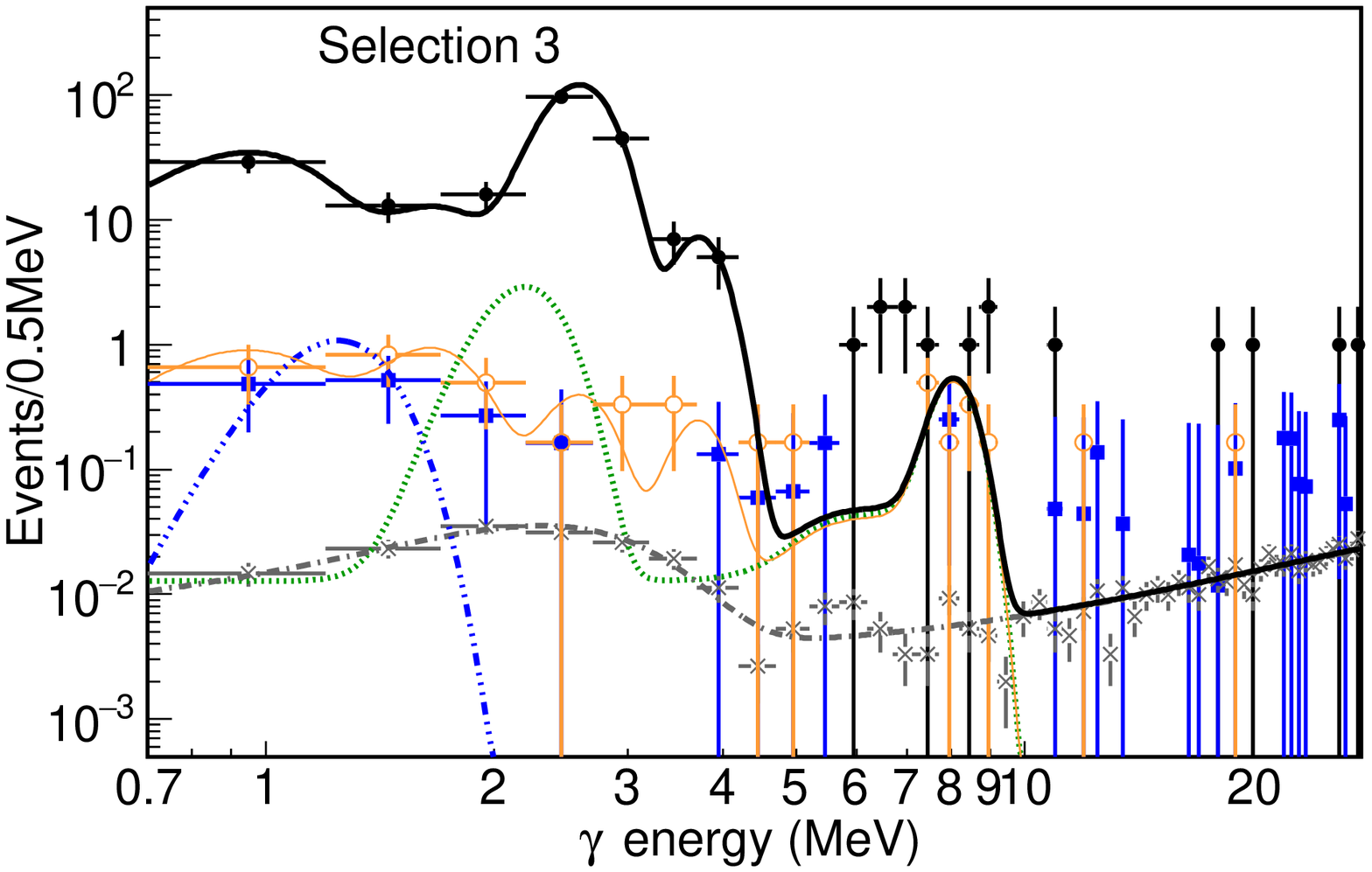}
\end{center}

\caption{\label{fig:b12gammafit} \is{12}B$\gamma$ spectra for each selection,
shown on a linear scale on the left and a log scale on the right. Selected
events and the overall fit are shown in black with filled circles and a thick
solid line, the signal components of the fit with a dashed black line,
accidental background (section~\ref{sec:gammaaccidentals}) in gray with crosses
and a dot-dashed line, correlated background from non-\is{12}B isotopes assumed
to be all \is{8}Li (section~\ref{sec:gammacorrbg}) in blue with squares and a
triple-dot-dashed line, and from early neutron captures following
\nrn{\ism{13}C}{n}{\ism{12}B} selected as gammas (section~\ref{sec:ncontam})
with a green dashed line.  The data and fit for
\nrn{\ism{13}C}{n}{\ism{12}B^{*}} (section~\ref{sec:gammab12n}) is shown in
orange with open circles and a thin solid line and is scaled to the level at
which it acts as a background to \nrn{\ism{12}C}{}{\ism{12}B^{*}}.  The fit
line includes the spallation background to \nrn{\ism{13}C}{n}{\ism{12}B} itself
(section~\ref{sec:gammathru}), visible at the \gdn energy, which does not
affect \nrn{\ism{12}C}{}{\ism{12}B}.} 

\end{figure*}}

\newcommand{\bgammacompfig}[1][tbp]{\begin{figure*}
\begin{center}
\includegraphics[width=1.75\columnwidth]{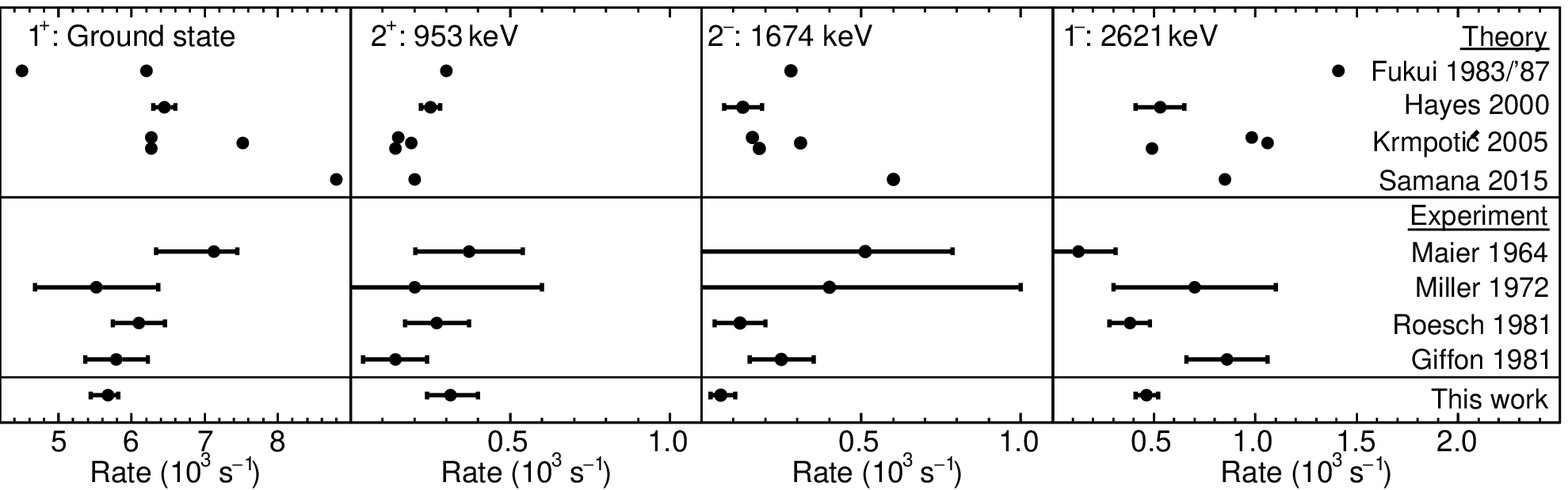}
\end{center}

\caption{\label{fig:b12gcomp} Comparison of results for each bound state of
\is{12}B to theory (from top to bottom: \cite{measday,fukui1,fukui2},
\cite{hayes}, \cite{krmpotic}, \cite{krm2}) and previous measurements
(\cite{maier}, \cite{miller}, \cite{roesch}, \cite{giffon}).
Krmpoti{\'c}~\cite{krmpotic} presents three parameterizations of PQRPA; all are
displayed here. Each of the more recent experiments depends on
Maier~\cite{maier} for the total rate, but here is shown updated using the
Double Chooz total rate.  Maier's result has itself been corrected here to
account for their \is{13}C fraction and to correct an error in their
calculations. Giffon~\cite{giffon} is reanalyzed to remove the erroneous
assumption that the 953\,keV level is negligible. One experiment~\cite{bud} has
been excluded due to having made the same assumption and there being
insufficient information to reanalyze.}

\end{figure*}}

\newcommand{\hesixtab}{
\begin{table}

\caption{\label{tab:he6} Results for \is{6}He search with various neutron
requirements.}

\setlength{\extrarowheight}{0.35ex}
\begin{center}
\begin{tabular}{c c c c}
\hline
\hline
$N_\mathrm{n}$ & signif. & limit & value \\
\hline
0 & 0.5$\sigma$ & $<0.43\%$ & --- \\
\hline
1 & 2.7$\sigma$ & $<0.47\%$ & $(0.33^{+0.14}_{-0.13}(\mathrm{stat})\pm0.02(\mathrm{syst}))$\% \\
\hline
2 & 0.5$\sigma$ & $<7\times10^{-4}$ & --- \\
\hline
3 & 0.2$\sigma$ & $<8\times10^{-4}$ & --- \\
\hline
\hline
\end{tabular}
\end{center}
\end{table}
}

\newcommand{\primaryresult}{$(2.4\pm0.9\mathrm{(stat)}\pm0.04\mathrm{(syst)})\times10^{-4}$}
\newcommand{\probNineLiFromThirteenC}{$(2.4\pm0.9\mathrm{(stat)}\pm0.05\mathrm{(syst)})\%$}

\newcommand{\probEightHeFromNatC}{$<7\times10^{-4}$}

\newcommand{\probElevenLiFromThirteenC}{$<0.7\%$}
\newcommand{\probTwelveBfromTwelveC}{$(17.35^{+0.21}_{-0.52}(\mathrm{stat})\pm0.27(\mathrm{syst}))$\%}
\newcommand{\probTwelveBFromThirteenC}{$(51.6\pm5.0(\mathrm{stat})\pm2.6(\mathrm{syst}))$\%}
\newcommand{\probThirteenBFromThirteenCbeta}{$<40\%$}
\newcommand{\probThirteenBFromThirteenCcombined}{\probThirteenBFromThirteenCbeta}
\newcommand{\linineheeightsignificance}{2.7$\sigma$}

\newcommand{\probEightLiFromTwelveC}{$(0.64\pm0.04\mathrm{(stat)}\pm0.02\mathrm{(syst)})\%$}
\newcommand{\probEightLiFromThirteenC}{$(5.1^{+1.1}_{-1.0}\mathrm{(stat)}\pm0.3\mathrm{(syst)})\%$}

\newcommand{\probEightBFromTwelveC}{$<3.2\times10^{-5}$}
\newcommand{\probTwelveNfromSixteenO}{$<8\times10^{-4}$}
\newcommand{\probNineCfromSixteenO}{$<9\times10^{-4}$}
\newcommand{\probNineCfromFourteenN}{$<3.6$\%}
\newcommand{\probElevenBefromTwelveC}{$<0.20\%$}
\newcommand{\probFifteenCfromSixteenO}{$<9\%$}
\newcommand{\probSixHeFromTwelveCZeron}{$<0.43\%$}
\newcommand{\probSixHeFromTwelveCnlim}{$<0.47\%$}

\newcommand{\probSixHeFromTwelveCnn}{$<7\times10^{-4}$}
\newcommand{\probSixHeFromTwelveCnnn}{$<8\times10^{-4}$}
\newcommand{\probTwelveBeFromThirteenC}{$<0.20\%$}
\newcommand{\probFourteenBfromSixteenO}{$<0.16\%$}

\newcommand{\probGammaZeroRateCent}{5.68}
\newcommand{\probGammaZeroRate}[1]
{$\probGammaZeroRateCent^{+0.13}_{-0.22}{\ifthenelse{#1 = 1}{\mathrm{(stat)}}{}}\pm0.06{\ifthenelse{#1 = 1}{\mathrm{(syst)}}{}}$}

\newcommand{\probGammaZeroRateSimpleWithVal}[1]
{$\probGammaZeroRateCent^{+0.14}_{-0.23}{\ifthenelse{#1 = 1}{\mathrm{(total)}}{}}$}

\newcommand{\probGammaZeroRateSimple}[1]
{$\phantom\probGammaZeroRateCent^{+0.14}_{-0.23}{\ifthenelse{#1 = 1}{\mathrm{(total)}}{}}$}

\newcommand{\probGammaOneFracCent}{4.8}
\newcommand{\probGammaOneFrac}[1]
{$\probGammaOneFracCent^{+1.3}_{-1.1}{\ifthenelse{#1 = 1}{\mathrm{(stat)}}{}}^{+0.2}_{-0.1}
                    \ifthenelse{#1 = 1}{\mathrm{(syst)}}{}$}
\newcommand{\probGammaOneFracSimple}[1]
{$\phantom\probGammaOneFracCent^{+1.3}_{-1.1}{\ifthenelse{#1 = 1}{\mathrm{(total)}}{}}$}
\newcommand{\probGammaOneRateCent}{0.31}
\newcommand{\probGammaOneRate}[1]
{$\probGammaOneRateCent^{+0.09}_{-0.07}{\ifthenelse{#1 = 1}{\mathrm{(stat)}}{}}\pm0.01
                      {\ifthenelse{#1 = 1}{\mathrm{(syst)}}{}}$}
\newcommand{\probGammaOneRateSimple}[1]
{$\phantom\probGammaOneRateCent^{+0.09}_{-0.07}{\ifthenelse{#1 = 1}{\mathrm{(total)}}{}}$}

\newcommand{\probGammaTwoFracCent}{0.9}
\newcommand{\probGammaTwoFrac}[1]
{$\probGammaTwoFracCent^{+0.7}_{-0.5}{\ifthenelse{#1 = 1}{\mathrm{(stat)}}{}}^{+0.05}_{-0.04}{\ifthenelse{#1 = 1}{\mathrm{(syst)}}{}}$}
\newcommand{\probGammaTwoFracSimple}[1]
{$\phantom\probGammaTwoFracCent^{+0.7}_{-0.5}{\ifthenelse{#1 = 1}{\mathrm{(total)}}{}}$}
\newcommand{\probGammaTwoRateCent}{0.06}
\newcommand{\probGammaTwoRate}[1]
{$\probGammaTwoRateCent^{+0.04}_{-0.03}{\ifthenelse{#1 = 1}{\mathrm{(stat)}}{}}\pm0.003{\ifthenelse{#1 = 1}{\mathrm{(syst)}}{}}$}
\newcommand{\probGammaTwoRateSimple}[1]
{$\phantom\probGammaTwoRateCent^{+0.04}_{-0.03}{\ifthenelse{#1 = 1}{\mathrm{(total)}}{}}$}

\newcommand{\probGammaThreeFracCent}{7.1}
\newcommand{\probGammaThreeFrac}[1]
{$\probGammaThreeFracCent^{+0.9}_{-0.8}{\ifthenelse{#1 = 1}{\mathrm{(stat)}}{}}^{+0.2}_{-0.1}{\ifthenelse{#1 = 1}{\mathrm{(syst)}}{}}$}
\newcommand{\probGammaThreeFracSimple}[1]
{$\phantom\probGammaThreeFracCent^{+0.9}_{-0.8}{\ifthenelse{#1 = 1}{\mathrm{(total)}}{}}$}
\newcommand{\probGammaThreeRateCent}{0.47}
\newcommand{\probGammaThreeRate}[1]
{$\probGammaThreeRateCent^{+0.06}_{-0.05}{\ifthenelse{#1 = 1}{\mathrm{(stat)}}{}}\pm0.01{\ifthenelse{#1 = 1}{\mathrm{(syst)}}{}}$}
\newcommand{\probGammaThreeRateSimple}[1]
{$\phantom\probGammaThreeRateCent^{+0.06}_{-0.05}{\ifthenelse{#1 = 1}{\mathrm{(total)}}{}}$}

\newcommand{\probGammaFourFracCent}{0.39}
\newcommand{\probGammaFourFrac}[1]
{$\probGammaFourFracCent^{+0.23}_{-0.17}{\ifthenelse{#1 = 1}{\mathrm{(stat)}}{}}\pm0.01{\ifthenelse{#1 = 1}{\mathrm{(syst)}}{}}$}
\newcommand{\probGammaFourFracSimple}[1]
{$\phantom\probGammaFourFracCent^{+0.23}_{-0.17}{\ifthenelse{#1 = 1}{\mathrm{(total)}}{}}$}
\newcommand{\probGammaFourRateCent}{0.026}
\newcommand{\probGammaFourRate}[1]
{$\probGammaFourRateCent^{+0.015}_{-0.011}{\ifthenelse{#1 = 1}{\mathrm{(stat)}}{}}\pm0.001{\ifthenelse{#1 = 1}{\mathrm{(syst)}}{}}$}
\newcommand{\probGammaFourRateSimple}[1]
{$\phantom\probGammaFourRateCent^{+0.015}_{-0.011}{\ifthenelse{#1 = 1}{\mathrm{(total)}}{}}$}

\newcommand{\GammaOneSig}  {$6.2\sigma$}
\newcommand{\GammaTwoSig}  {$1.6\sigma$}
\newcommand{\GammaThreeSig}{$15.3\sigma$}
\newcommand{\GammaFourSig} {$2.8\sigma$}

\newcommand{\nprobGammaZeroRateCent}{X}
\newcommand{\nprobGammaZeroRate}[1]
{$\nprobGammaZeroRateCent^{X}_{X}{\ifthenelse{#1 = 1}{\mathrm{(stat)}}{}}\pm X{\ifthenelse{#1 = 1}{\mathrm{(syst)}}{}}$}

\newcommand{\nprobGammaZeroRateSimple}[1]
{$\nprobGammaZeroRateCent^{X}_{X}{\ifthenelse{#1 = 1}{\mathrm{(total)}}{}}$}

\newcommand{\nprobGammaOneFracCent}{46}
\newcommand{\nprobGammaOneFrac}[1]
{$\nprobGammaOneFracCent^{+29}_{-21}{\ifthenelse{#1 = 1}{\mathrm{(stat)}}{}}\pm4
                    \ifthenelse{#1 = 1}{\mathrm{(syst)}}{}$}
\newcommand{\nprobGammaOneFracSimple}[1]
{$<80.$ at 90\% CL}
\newcommand{\nprobGammaOneRateCent}{8}
\newcommand{\nprobGammaOneRate}[1]
{$\nprobGammaOneRateCent^{+5}_{-4}{\ifthenelse{#1 = 1}{\mathrm{(stat)}}{}}\pm0.2
                      {\ifthenelse{#1 = 1}{\mathrm{(syst)}}{}}$}
\newcommand{\nprobGammaOneRateSimple}[1]
{$<15$ at 90\% CL}

\newcommand{\nprobGammaTwoFracCent}{25}
\newcommand{\nprobGammaTwoFrac}[1]
{$\nprobGammaTwoFracCent^{+22}_{-14}{\ifthenelse{#1 = 1}{\mathrm{(stat)}}{}}\pm 2{\ifthenelse{#1 = 1}{\mathrm{(syst)}}{}}$}
\newcommand{\nprobGammaTwoFracSimple}[1]
{$<54$ at 90\% CL}
\newcommand{\nprobGammaTwoRateCent}{5}
\newcommand{\nprobGammaTwoRate}[1]
{$\nprobGammaTwoRateCent^{+4}_{-3}{\ifthenelse{#1 = 1}{\mathrm{(stat)}}{}}\pm0.1{\ifthenelse{#1 = 1}{\mathrm{(syst)}}{}}$}
\newcommand{\nprobGammaTwoRateSimple}[1]
{$<9.9$ at 90\% CL}

\newcommand{\nprobGammaThreeFracCent}{4}
\newcommand{\nprobGammaThreeFrac}[1]
{$\nprobGammaThreeFracCent^{+6}_{-3}{\ifthenelse{#1 = 1}{\mathrm{(stat)}}{}}\pm 0.3{\ifthenelse{#1 = 1}{\mathrm{(syst)}}{}}$}
\newcommand{\nprobGammaThreeFracSimple}[1]
{$<12$ at 90\% CL}
\newcommand{\nprobGammaThreeRateCent}{0.7}
\newcommand{\nprobGammaThreeRate}[1]
{$\nprobGammaThreeRateCent^{+1.1}_{-0.6}{\ifthenelse{#1 = 1}{\mathrm{(stat)}}{}}\pm0.02{\ifthenelse{#1 = 1}{\mathrm{(syst)}}{}}$}
\newcommand{\nprobGammaThreeRateSimple}[1]
{$<2.2$ at 90\% CL}

\newcommand{\nprobGammaFourFracCent}{4}
\newcommand{\nprobGammaFourFrac}[1]
{$\nprobGammaFourFracCent^{+6}_{-3}{\ifthenelse{#1 = 1}{\mathrm{(stat)}}{}}\pm0.4{\ifthenelse{#1 = 1}{\mathrm{(syst)}}{}}$}
\newcommand{\nprobGammaFourFracSimple}[1]
{$<12$ at 90\% CL}
\newcommand{\nprobGammaFourRateCent}{0.8}
\newcommand{\nprobGammaFourRate}[1]
{$\nprobGammaFourRateCent^{+1.1}_{-0.6}{\ifthenelse{#1 = 1}{\mathrm{(stat)}}{}}\pm0.02{\ifthenelse{#1 = 1}{\mathrm{(syst)}}{}}$}
\newcommand{\nprobGammaFourRateSimple}[1]
{$<2.2$ at 90\% CL}

\newcommand{\nGammaOneSig}  {$3.0\sigma$}
\newcommand{\nGammaTwoSig}  {$1.6\sigma$}
\newcommand{\nGammaThreeSig}{$1.2\sigma$}
\newcommand{\nGammaFourSig} {$1.8\sigma$}

\newcommand{\probLiEightGammaFracCent}{40}
\newcommand{\probLiEightGammaFrac}[1]
{$\probLiEightGammaFracCent \pm 20{\ifthenelse{#1 = 1}{\mathrm{(stat)}}{}}\pm3{\ifthenelse{#1 = 1}{\mathrm{(syst)}}{}}$}
\newcommand{\probLiEightGammaFracSimple}[1]{$<67\%$ at 90\% CL}

\newcommand{\probLiEightGammaRateCent}{0.10}
\newcommand{\probLiEightGammaRate}[1]
{$\probLiEightGammaRateCent \pm 0.05{\ifthenelse{#1 = 1}{\mathrm{(stat)}}{}}\pm0.002{\ifthenelse{#1 = 1}{\mathrm{(syst)}}{}}$}
\newcommand{\probLiEightGammaRateSimple}[1]
{$<0.16$ at 90\% CL}

\newcommand{\summarytabletable}{
\begin{table}

\caption{\label{tab:summary} All measurements of isotope production
probabilities per nuclear muon capture in this work. All limits are set at 90\%
CL and include systematic errors. Protons and alphas are shown as products for
illustration; they are not observed. The reaction product ``X'' indicates
several possible unobserved additional products.  The \is{9}Li results assume
no \is{8}He, and the \nrn{\ism{13}C}{\alpha}{\ism{9}Li} result assumes no
contribution from \is{12}C. }

\setlength{\extrarowheight}{0.35ex}
\begin{tabular}{r@{ }l c}
\hline
\hline
\multicolumn{2}{c}{Reaction}  & Probability/capture \\ \hline
\hline
\trn{\ism{12}C}{0nX}{\ism{6}He} & \probSixHeFromTwelveCZeron \\ \hline
\trn{\ism{12}C}{1nX}{\ism{6}He} & \probSixHeFromTwelveCnlim \\ \hline
\trn{\ism{12}C}{2nX}{\ism{6}He} & \probSixHeFromTwelveCnn \\ \hline
\trn{\ism{12}C}{3nX}{\ism{6}He} & \probSixHeFromTwelveCnnn \\ \hline
\trn{\ism{12}C}{4n}{\ism{8}B} & \probEightBFromTwelveC \\ \hline
\trn{\ism{12}C}{\alpha}{\ism{8}Li} & \probEightLiFromTwelveC  \\ \hline
\trn{\ism{12}C}{p}{\ism{11}Be} & \probElevenBefromTwelveC \\ \hline
\trn{\ism{12}C}{}{\ism{12}B}  & \probTwelveBfromTwelveC \\ \hline
\hline
\trn{\ism{\mathrm{nat}}C}{X}{\ism{8}He} & \probEightHeFromNatC \\ \hline
\trn{\ism{\mathrm{nat}}C}{X}{\ism{9}Li} & \primaryresult \\
\hline
\hline
\trn{\ism{13}C}{n\alpha}{\ism{8}Li} & \probEightLiFromThirteenC \\ \hline
\trn{\ism{13}C}{\alpha}{\ism{9}Li} & \probNineLiFromThirteenC \\ \hline
\trn{\ism{13}C}{2p}{\ism{11}Li} & \probElevenLiFromThirteenC \\ \hline
\trn{\ism{13}C}{p}{\ism{12}Be} & \probTwelveBeFromThirteenC \\ \hline
\trn{\ism{13}C}{n}{\ism{12}B} & \probTwelveBFromThirteenC \\ \hline
\trn{\ism{13}C}{}{\ism{13}B}  & \probThirteenBFromThirteenCcombined \\ \hline
\hline
\trn{\ism{14}N}{5n}{\ism{9}C} & \probNineCfromFourteenN \\ \hline
\hline
\trn{\ism{16}O}{2p}{\ism{14}B} & \probFourteenBfromSixteenO \\ \hline
\trn{\ism{16}O}{p6n}{\ism{9}C} & \probNineCfromSixteenO \\ \hline
\trn{\ism{16}O}{p}{\ism{15}C} & \probFifteenCfromSixteenO \\ \hline
\trn{\ism{16}O}{4n}{\ism{12}N} & \probTwelveNfromSixteenO \\ \hline
\hline
\end{tabular}
\end{table}
}

\newcommand{\isotable}{
\begin{table}[h]

\caption{\label{tab:iso} Beta-decaying isotopes that can be produced by muon
capture in Double Chooz. The lower left box contains those that can be produced
from \is{12}C.}

\begin{center}
\setlength{\extrarowheight}{0.35ex}
\noindent\begin{tabular}
{ c c c c c | c c c c c}
\fast{10}{N}&\fast{11}{N}&\isok{12}{N}&\slow{13}{N} &\stab{14}{N} &\stab{15}{N }&\isok{16}{N}&\rare{17}{N} &\rare{18}{N} \\
\isok{9}{C} &\slow{10}{C}&\slow{11}{C}&\stab{12}{C} &\stab{13}{C} &\slow{14}{C }&\isok{15}{C}&\rare{16}{C} &\rare{17}{C} \\
\hline
\isok{8}{B}& \fast{9}{B} &\stab{10}{B}&\stab{11}{B} &\isok{12}{B} &\isok{13}{B }&\isok{14}{B}&\rare{15}{B} &~            \\
\slow{7}{Be}&\fast{8}{Be}&\stab{9}{Be}&\slow{10}{Be}&\isok{11}{Be}&\isok{12}{Be}&~           &\rare{14}{Be}&~            \\
\stab{6}{Li}&\stab{7}{Li}&\isok{8}{Li}&\isok{9}{Li} &\fast{10}{Li}&\isok{11}{Li}&~           &~            &~            \\
\fast{5}{He}&\isok{6}{He}&\fast{7}{He}&\isok{8}{He} &\fast{9}{He} &~            &~           &~            &~            \\
\end{tabular}
\end{center}
\end{table}}

\newcommand{\isotablemore}{
\begin{table*}

\caption{\label{tab:isomore} Decay characteristics~\cite{nndc} for all
detectable isotopes that could be produced, in principle, from muon capture in
Double Chooz, and the most likely reaction(s) producing them.  Reactions are 
marked as ``exotic'' if more than one charged particle or more
than two neutrons are emitted, and ``obs.'' if they have been observed prior
to this work.}

\begin{center}
\setlength{\extrarowheight}{0.35ex}
\noindent\begin{tabular}
{ c c d d d r@{ }l c }
\hline
\hline
Isotope & Decay mode & \multicolumn{1}{c}{Half-life (ms)} &
\multicolumn{1}{c}{$Q_\beta$ (MeV)} & \multicolumn{1}{c}{$Q_{\beta\mathrm n}$} &
\multicolumn{2}{c}{Reaction} & Comment \\
\hline
\is{6}He  & \betam                   &      801    &  3.505 &      ~ & \trn{\ism{12}C}{np\alpha}{\ism{6}He} & Exotic \\
\hline
\is{8}He  & \betam: 16\% \betan      &      119.1  & 10.664 &  8.631 & \trn{\ism{13}C}{p\alpha}{\ism{8}He}  & Exotic \\
\hline
\is{8}Li  & $\beta^-\alpha$          &      839.9  & 16.004 &      ~ & \trn{\ism{12}C}{\alpha}{\ism{8}Li}   &        \\
          &                          &             &        &        & \trn{\ism{13}C}{n\alpha}{\ism{8}Li}  &        \\
\hline
\is{9}Li  & \betam: 50.8\% \betan    &      178.3  & 13.606 & 11.942 & \trn{\ism{13}C}{\alpha}{\ism{9}Li}   &        \\
\hline
\is{11}Li & \betam: 83\% \betan      &        8.75 & 20.551 & 20.049 & \trn{\ism{13}C}{2p}{\ism{11}Li}      & Exotic \\
\hline
\is{11}Be & \betam                   & 13\,810     & 11.509 &      ~ & \trn{\ism{12}C}{p}{\ism{11}Be}       &        \\
\hline
\is{12}Be & \betam                   &      21.49  & 11.708 &      ~ & \trn{\ism{13}C}{p}{\ism{12}Be}       &        \\
\hline
\is{14}Be & \betam: 81\% \betan      &       4.35  & 16.29  & 15.32  & \trn{\ism{18}O}{n3p}{\ism{14}Be}     & Exotic \\
\hline
\is{8}B   & $\beta^+\alpha$          &     770     & 16.958 &      ~ & \trn{\ism{12}C}{4n}{\ism{8}B}        & Exotic \\
\hline
\is{12}B  & \betam                   &      20.20  & 13.369 &      ~ & \trn{\ism{12}C}{}{\ism{12}B}         & Obs. \\
          &                          &             &        &        & \trn{\ism{13}C}{n}{\ism{12}B}        &       \\
\hline
\is{13}B  & \betam: 0.29\% \betan    &      17.33  & 13.437 &  8.491 & \trn{\ism{13}C}{}{\ism{13}B}         &        \\
\hline
\is{14}B  & \betam                   &      12.5   & 20.644 &      ~ & \trn{\ism{16}O}{2p}{\ism{14}B}       & Exotic \\
\hline
\is{15}B  & \betam: 93.6\% \betan    &       9.93  & 19.085 & 17.867 & \trn{\ism{18}O}{n2p}{\ism{15}B}      & Exotic \\
\hline
\is{9}C   & $\beta^+\mathrm p\alpha$ &     126.5   & 15.473 &      ~ & \trn{\ism{14}N}{5n}{\ism{9}C}        & Exotic \\
\hline
\is{15}C  & \betam                   &    2449     &  9.772 &      ~ & \trn{\ism{16}O}{p}{\ism{15}C}        & Exotic \\
\hline
\is{16}C  & \betam: 99\% \betan      &     747     &  8.010 &  5.521 & \trn{\ism{18}O}{np}{\ism{16}C}       &        \\
\hline
\is{17}C  & \betam: 32\% \betan      &     193     & 13.161 &  7.276 & \trn{\ism{18}O}{p}{\ism{17}C}        &        \\
\hline
\is{12}N  & \betap                   &      11.000 & 16.316 &      ~ & \trn{\ism{16}O}{4n}{\ism{12}N}       &        \\
\hline
\is{16}N  & \betam                   &    7130     & 10.421 &      ~ & \trn{\ism{16}O}{}{\ism{16}N}         & Obs.  \\
\hline
\is{17}N  & \betam: 95.1\% \betan    &    4173     &  8.679 &  4.536 & \trn{\ism{18}O}{n}{\ism{17}N}        &        \\
\hline
\is{18}N  & \betam                   &     620     & 13.896 &      ~ & \trn{\ism{18}O}{}{\ism{18}N}         &        \\
\hline
\hline

\end{tabular}
\end{center}
\end{table*}}

\newcommand{\borontab}{
\begin{table}

\caption{\label{tab:boron} Results for \is{12}B and \is{13}B production from
carbon, expressed as a probability per atomic capture (top), per nuclear
capture (middle) and as a transition rate (bottom).  In each case, the total
error is given, and then each component of the error is shown.  $N_{\mu^-}$
gives the error due to \mum counting, P(cap) is the error due to knowledge of
the total nuclear capture rate, and $\tau_{\mu^-}$ is the error due to the
knowledge of the bound \mum lifetime.  Limits are given at 90\% CL.  }

\setlength{\extrarowheight}{0.35ex}
\begin{tabular}{c c c c c}
\hline
\hline
\multicolumn{5}{c}{Probability per atomic capture (\%)} \\
\hline
Reaction & Prob., total error & Stat & $N_{\mu^-}$ & \is{12}B eff. \\
\hline
\is{12}C$\rightarrow$\is{12}B &
$1.322^{+0.022}_{-0.042}$ &
$^{+0.016}_{-0.040}$ &
$\pm0.010$ &
$\pm0.010$ \\

\is{13}C$\rightarrow$\is{12}Bn &
$3.75\pm0.36$ &
$\pm0.36$ &
$\pm0.03$ &
$\pm0.03$ \\

\is{13}C$\rightarrow$\is{13}B &
$<2.8$ &
&
&
\\
\hline
\hline
\end{tabular}

\vspace{2ex}

\begin{tabular}{c c c c c c}
\hline
\hline
\multicolumn{6}{c}{Probability per nuclear capture (\%)} \\
\hline
Reaction & Prob., total error & Stat & $N_{\mu^-}$ & \is{12}B eff. & P(cap) \\
\hline
\is{12}C$\rightarrow$\is{12}B &
$17.35^{+0.35}_{-0.59}$ &
$^{+0.21}_{-0.52}$ &
$\pm0.13$ &
$\pm0.13$ &
$\pm0.21$ \\

\is{13}C$\rightarrow$\is{12}Bn &
$51.6\pm5.6$ &
$\pm5.0$ &
$\pm0.4$ &
$\pm0.4$ &
$\pm2.6$ \\

\is{13}C$\rightarrow$\is{13}B &
$<40$ &
&
&
&
\\
\hline
\hline
\end{tabular}

\vspace{2ex}

\begin{tabular}{c c c c c c}
\hline
\hline
\multicolumn{6}{c}{Capture rate ($\times10^{3}\mathrm{s}^{-1}$)} \\
\hline
Reaction & Rate, total error & Stat & $N_{\mu^-}$ & \is{12}B eff. & $\tau_{\mu^-}$ \\
\hline
\is{12}C$\rightarrow$\is{12}B &
$6.57^{+0.11}_{-0.21}$ &
$^{+0.08}_{-0.20}$ &
$\pm0.05$ &
$\pm0.05$ &
$\pm0.01$ \\

\is{13}C$\rightarrow$\is{12}Bn &
$18.4\pm1.8$ &
$\pm1.8$ &
$\pm0.1$ &
$\pm0.1$ &
$\pm0.1$ \\

\is{13}C$\rightarrow$\is{13}B &
$<14$ &
&
&
&
\\
\hline
\hline
\end{tabular}
\end{table}}

\newcommand{\gammaeightsummarytabletable}{
\setlength{\extrarowheight}{0.6ex}
\begin{tabular}{l l l}
\hline
\hline
Daughter & Rate ($10^3\,\mathrm s^{-1})$ & Per \is{8}Li $\beta$ decay (\%) \\ \hline
\hline
\is{8}Li*(981)& \probLiEightGammaRate{1}             & \probLiEightGammaFrac{1}             \\
              & \ba{\probLiEightGammaRateSimple{1}}  & \ba{\probLiEightGammaFracSimple{1}}  \\ \hline
\hline
\end{tabular}
}

\newcommand{\gammathirteensummarytabletable}{
\setlength{\extrarowheight}{0.6ex}
\begin{tabular}{l l l c}
\hline
\hline
Daughter & Rate ($10^3\,\mathrm s^{-1})$ & Per \is{12}B $\beta$ decay (\%) & Significance \\ \hline
\hline
\is{12}B*(953)+n & \nprobGammaOneRate{1}              & \nprobGammaOneFrac{1}              & \nGammaOneSig   \\
                 & \ba{\nprobGammaOneRateSimple{1}}   & \ba{\nprobGammaOneFracSimple{1}}   &                 \\ \hline
\is{12}B*(1674)+n& \nprobGammaTwoRate{1}              & \nprobGammaTwoFrac{1}              & \nGammaTwoSig   \\
                 & \ba{\nprobGammaTwoRateSimple{1}}   & \ba{\nprobGammaTwoFracSimple{1}}   &                 \\ \hline
\is{12}B*(2621)+n& \nprobGammaThreeRate{1}            & \nprobGammaThreeFrac{1}            & \nGammaThreeSig \\
                 & \ba{\nprobGammaThreeRateSimple{1}} & \ba{\nprobGammaThreeFracSimple{1}} &                 \\ \hline
\is{12}B*(3759)+n& \nprobGammaFourRate{1}             & \nprobGammaFourFrac{1}             & \nGammaFourSig  \\
                 & \ba{\nprobGammaFourRateSimple{1}}  & \ba{\nprobGammaFourFracSimple{1}}  &                 \\ \hline
\hline
\end{tabular}
}

\newcommand{\gammaeightsummarytable}{
\begin{table*}

\caption{\label{tab:gammaeightsummary} Rate for the exclusive reaction
\is{12}C$(\mu^-,\nu\alpha)^8\mathrm{Li}^*(981)$.  The methodology is the same
as in table~\ref{tab:gammathirteensummary}.}

\begin{center}
\gammaeightsummarytabletable
\end{center}
\end{table*}}

\newcommand{\gammathirteensummarytable}{
\begin{table*}

\caption{\label{tab:gammathirteensummary} Transition rates for each \is{12}B
level measured in the reactions \is{13}C$(\mu^-,\nu\mathrm n)$\is{12}B$^{*}$.
The limits are taken as the primary results. They result from a Bayesian
analysis that takes a flat prior from 0--100\%. Because it is assumed that any
or all of the events are background, none of the levels are imposed as
constraints on the probabilities of other levels.}

\begin{center}
\gammathirteensummarytabletable
\end{center}
\end{table*}}

\newcommand{\gammasummarytabletable}{
\setlength{\extrarowheight}{0.6ex}
\begin{tabular}{l l l c}
\hline
\hline
Daughter & Rate ($10^3\,\mathrm s^{-1})$ & Per \is{12}B $\beta$ decay (\%) & Significance \\ \hline
\hline
\is{12}B(g.s.)   & \probGammaZeroRate{1}        &                              &                \\
                 & \probGammaZeroRateSimple{1}  &                              &                \\ \hline
\is{12}B*(953)   & \probGammaOneRate{1}         & \probGammaOneFrac{1}         & \GammaOneSig   \\
                 & \probGammaOneRateSimple{1}   & \probGammaOneFracSimple{1}   &                \\ \hline
\is{12}B*(1674)  & \probGammaTwoRate{1}         & \probGammaTwoFrac{1}         & \GammaTwoSig   \\
                 & \probGammaTwoRateSimple{1}   & \probGammaTwoFracSimple{1}   &                \\ \hline
\is{12}B*(2621)  & \probGammaThreeRate{1}       & \probGammaThreeFrac{1}       & \GammaThreeSig \\
                 & \probGammaThreeRateSimple{1} & \probGammaThreeFracSimple{1} &                \\ \hline
\is{12}B*(3759)  & \probGammaFourRate{1}        & \probGammaFourFrac{1}        & \GammaFourSig  \\
                 & \probGammaFourRateSimple{1}  & \probGammaFourFracSimple{1}  &                \\ \hline
\hline
\end{tabular}
}

\newcommand{\gammasummarytable}{
\begin{table*}

\caption{\label{tab:gammasummary} Transition rates to each \is{12}B level
measured in the reactions \is{12}C$(\mu^-,\nu)$\is{12}B$^{(*)}$. For 3759\,keV,
this is the probability of reaching the level \emph{and} de-exciting via gamma
emission; it is assumed that no other gamma lines contribute. }

\begin{center}
\gammasummarytabletable
\end{center}
\end{table*}}

\begin{document}

\title{Muon capture on light isotopes in Double Chooz}

\newcommand\Aachen{III. Physikalisches Institut, RWTH Aachen University, 52056 Aachen, Germany}\affiliation{\Aachen}
\newcommand\Alabama{Department of Physics and Astronomy, University of Alabama, Tuscaloosa, Alabama 35487, USA}\affiliation{\Alabama}
\newcommand\Argonne{Argonne National Laboratory, Argonne, Illinois 60439, USA}\affiliation{\Argonne}
\newcommand\APC{AstroParticule et Cosmologie, Universit\'{e} Paris Diderot, CNRS/IN2P3, CEA/IRFU, Observatoire de Paris, Sorbonne Paris Cit\'{e}, 75205 Paris Cedex 13, France}\affiliation{\APC}
\newcommand\CBPF{Centro Brasileiro de Pesquisas F\'{i}sicas, Rio de Janeiro, RJ, 22290-180, Brazil}\affiliation{\CBPF}
\newcommand\Chicago{The Enrico Fermi Institute, The University of Chicago, Chicago, Illinois 60637, USA}\affiliation{\Chicago}
\newcommand\CIEMAT{Centro de Investigaciones Energ\'{e}ticas, Medioambientales y Tecnol\'{o}gicas, CIEMAT, 28040, Madrid, Spain}\affiliation{\CIEMAT}
\newcommand\Columbia{Columbia University; New York, New York 10027, USA}\affiliation{\Columbia}
\newcommand\Davis{University of California, Davis, California 95616, USA}\affiliation{\Davis}
\newcommand\Drexel{Department of Physics, Drexel University, Philadelphia, Pennsylvania 19104, USA}\affiliation{\Drexel}
\newcommand\Hiroshima{Hiroshima Institute of Technology, Hiroshima, 731-5193, Japan}\affiliation{\Hiroshima}
\newcommand\IIT{Department of Physics, Illinois Institute of Technology, Chicago, Illinois 60616, USA}\affiliation{\IIT}
\newcommand\INR{Institute of Nuclear Research of the Russian Academy of Sciences, Moscow 117312, Russia}\affiliation{\INR}
\newcommand\CEA{Commissariat \`{a} l'Energie Atomique et aux Energies Alternatives, Centre de Saclay, IRFU, 91191 Gif-sur-Yvette, France}\affiliation{\CEA}
\newcommand\Kansas{Department of Physics, Kansas State University, Manhattan, Kansas 66506, USA}\affiliation{\Kansas}
\newcommand\Kitasato{Department of Physics, Kitasato University, Sagamihara, 252-0373, Japan}\affiliation{\Kitasato}
\newcommand\Kobe{Department of Physics, Kobe University, Kobe, 657-8501, Japan}\affiliation{\Kobe}
\newcommand\Kurchatov{NRC Kurchatov Institute, 123182 Moscow, Russia}\affiliation{\Kurchatov}
\newcommand\MIT{Massachusetts Institute of Technology, Cambridge, Massachusetts 02139, USA}\affiliation{\MIT}
\newcommand\MaxPlanck{Max-Planck-Institut f\"{u}r Kernphysik, 69117 Heidelberg, Germany}\affiliation{\MaxPlanck}
\newcommand\NotreDame{University of Notre Dame, Notre Dame, Indiana 46556, USA}\affiliation{\NotreDame}
\newcommand\IPHC{IPHC, Universit\'{e} de Strasbourg, CNRS/IN2P3, 67037 Strasbourg, France}\affiliation{\IPHC}
\newcommand\SUBATECH{SUBATECH, CNRS/IN2P3, Universit\'{e} de Nantes, Ecole des Mines de Nantes, 44307 Nantes, France}\affiliation{\SUBATECH}
\newcommand\Tennessee{Department of Physics and Astronomy, University of Tennessee, Knoxville, Tennessee 37996, USA}\affiliation{\Tennessee}
\newcommand\TohokuUni{Research Center for Neutrino Science, Tohoku University, Sendai 980-8578, Japan}\affiliation{\TohokuUni}
\newcommand\TohokuGakuin{Tohoku Gakuin University, Sendai, 981-3193, Japan}\affiliation{\TohokuGakuin}
\newcommand\TokyoInst{Department of Physics, Tokyo Institute of Technology, Tokyo, 152-8551, Japan }\affiliation{\TokyoInst}
\newcommand\TokyoMet{Department of Physics, Tokyo Metropolitan University, Tokyo, 192-0397, Japan}\affiliation{\TokyoMet}
\newcommand\Muenchen{Physik Department, Technische Universit\"{a}t M\"{u}nchen, 85748 Garching, Germany}\affiliation{\Muenchen}
\newcommand\Tubingen{Kepler Center for Astro and Particle Physics, Universit\"{a}t T\"{u}bingen, 72076 T\"{u}bingen, Germany}\affiliation{\Tubingen}
\newcommand\UFABC{Universidade Federal do ABC, UFABC, Santo Andr\'{e}, SP, 09210-580, Brazil}\affiliation{\UFABC}
\newcommand\UNICAMP{Universidade Estadual de Campinas-UNICAMP, Campinas, SP, 13083-970, Brazil}\affiliation{\UNICAMP}
\newcommand\vtech{Center for Neutrino Physics, Virginia Tech, Blacksburg, Virginia 24061, USA}\affiliation{\vtech}

\newcommand{\nowathawaii}{\altaffiliation[Now at ]{Department of Physics \& Astronomy, University of Hawaii at Manoa, Honolulu, Hawaii 96822, USA.}}
\newcommand{\nowatific}{\altaffiliation[Now at ]{Instituto de F\'{i}sica Corpuscular, IFIC (CSIC/UV), 46980 Paterna, Spain.}}
\newcommand{\nowatmaryland}{\altaffiliation[Now at ]{Department of Physics, University of Maryland, College Park, Maryland 20742, USA.}}
\newcommand{\nowatkobe}{\altaffiliation[Now at ]{Department of Physics, Kobe University, Kobe, 658-8501, Japan.}}
\newcommand{\nowatprisma}{\altaffiliation[Now at ]{Institut f\"{u}r Physik and Excellence Cluster PRISMA, Johannes Gutenberg-Universit\"{a}t Mainz, 55128 Mainz, Germany.}}
\newcommand{\nowatelinp}{\altaffiliation[Now at ]{ELI-NP, ``Horia Hulubei'' National Institute of Physics and Nuclear Engineering, 077125
Bucharest-Magurele, Romania}}

\author{Y.~Abe} \affiliation{\TokyoInst} 
\author{T.~Abrah\~{a}o} \affiliation{\CBPF} 
\author{H.~Almazan} \affiliation{\MaxPlanck} 
\author{C.~Alt} \affiliation{\Aachen} 
\author{S.~Appel} \affiliation{\Muenchen} 
\author{J.C.~Barriere} \affiliation{\CEA} 
\author{E.~Baussan} \affiliation{\IPHC} 
\author{I.~Bekman} \affiliation{\Aachen} 
\author{M.~Bergevin} \affiliation{\Davis} 
\author{T.J.C.~Bezerra} \affiliation{\TohokuUni} 
% ...

\collaboration{Double Chooz Collaboration}

\date{\today}

\begin{abstract} 

Using the Double Chooz detector, designed to measure the neutrino mixing angle
$\theta_{13}$, the products of \mum capture on \is{12}C, \is{13}C, \is{14}N and
\is{16}O have been measured.  Over a period of 489.5 days, $2.3\times10^6$
stopping cosmic \mum have been collected, of which $1.8\times10^5$ captured on
carbon, nitrogen, or oxygen nuclei in the \id scintillator or acrylic vessels.
The resulting isotopes were tagged using prompt neutron emission (when
applicable), the subsequent beta decays, and, in some cases,
\mbox{$\beta$-delayed} neutrons.  The most precise measurement of the rate of
\nrn{\ism{12}C}{}{\ism{12}B} to date is reported:
$6.57^{+0.11}_{-0.21}\times10^{3}\,\mathrm s^{-1}$, or
$(17.35^{+0.35}_{-0.59})\%$ of nuclear captures.  By tagging excited states
emitting gammas, the ground state transition rate to \is{12}{B} has been
determined to be \probGammaZeroRateSimpleWithVal{0}$\times10^3\,\mathrm s^{-1}$.
The heretofore unobserved reactions \nrn{\ism{12}C}{\alpha}{\ism{8}Li},
\nrn{\ism{13}C}{n\alpha}{\ism{8}Li}, and \nrn{\ism{13}C}{n}{\ism{12}B} are
measured.  Further, a population of \betan decays following stopping muons is
identified with 5.5$\sigma$ significance.  Statistics limit our ability to
identify these decays definitively. Assuming negligible production of \is{8}He,
the reaction \nrn{\ism{13}C}{\alpha}{\ism{9}Li} is found to be present at the
\linineheeightsignificance\ level. Limits are set on a variety of other
processes.

\end{abstract}

\maketitle


\section{Introduction}

A \mum that stops in a material is rapidly captured into an atomic orbital and
subsequently rapidly cascades down into the 1s state (referred to herein as
\emph{atomic capture}).  From this state, a muon can either decay or may
undergo \emph{nuclear capture}.  This process is the muon analog to the
electron capture decay that proton-rich isotopes undergo.  However, muon
capture is always energetically favorable and therefore competes with muon
decay regardless of the isotope.  The large mass of the muon also makes
available a wide variety of final states.  In addition to converting a proton
to a neutron, nucleons can be ejected from the nucleus~\cite{measday}.

Measurements of the probabilities of these various states are interesting for
reactor neutrino experiments, for example, because they form backgrounds to the
inverse beta decay reaction.  In particular, production of isotopes that
undergo beta-delayed neutron emission (henceforth called \betan isotopes) forms
a nearly irreducible background, since the inverse beta decay signal is a
positron and a neutron.  Production of ordinary beta-unstable isotopes is less
troublesome for reactor neutrino experiments, but as these isotopes often have
$Q_\beta$ values far in excess of natural radioactivity, they will contribute
to the high-energy accidental spectrum of delayed-coincidence searches. In
addition, it could be relevant for experiments whose signal is a single event,
for instance electron recoil. 

Muon capture measurements provide input for understanding nuclear structure and
can be used to validate models of neutrino cross sections, such as PQRPA.  This
is particularly true for measurements of production of the exclusive states of
\is{12}{B}~\cite{krmpotic}.  The reaction is a close analogue to neutrino
capture interactions such as those that will be observed in the event of a
galactic supernova.  In each, about $\sim 20\,$MeV is available for
excitations, often leading to nucleon emission and a variety of excited states.
Another similar situation occurs in the baryon-number violating process of
proton or bound-neutron decay to invisible products, as is being studied by
several large underground detectors. While not currently a competitive technique, the rate of
\is{12}{C}$(\mu^-,\nu)^{12}\mathrm{B_{g.s.}}$ has also been used as input for
measurement of $g_{P}$~\cite{gpreview}. 

The structure of this paper is as follows.  In \sect{sec:detector}, the
detector is described, including its isotopic composition.  In
\sect{sec:rateofstoppingmu}, the rate of stopping muons in a clean sample of
events (the ``high-purity sample'') is found.  This rate is used in
\sect{sec:b12analysis} to find the probabilities of forming \is{12}B and
\is{13}B.  These probabilities are then used in \sect{sec:ratesloose} to find
the stopping muon rate in a less clean, but three times larger, sample of
stopping muons (the ``loose sample'').  Total nuclear and atomic capture rates
for each relevant isotope, along with their systematic errors, are tabulated
here.  This larger sample is used to analyze production of isotopes in the
following sections.

Analysis of all the rest of the isotopes follows: \betan emitters are presented
in \sect{sec:betan}, the \is{12}Be decay chain in \sect{sec:decaychain}, and a
variety of simple beta decays in \sect{sec:beta}.  Section~\ref{sec:b12gamma}
shows exclusive measurements of \is{12}B states. Finally, \sect{sec:conclusion}
concludes.

\section{The Detector}\label{sec:detector}

The Double Chooz experiment consists of two nearly identical detectors, the
\emph{near} and \emph{far} detectors, placed 400\,m and 1\,km, respectively,
from a commercial nuclear reactor.  It has been described in detail
elsewhere~\cite{3rdpub}.  It consists of four concentric liquid zones, passive
shielding and an \ov of plastic scintillator.  The innermost three zones form
one optical volume as follows:

\ul

  \item The \emph{\ntns} is a cylinder with height 2.5\,m 
  and diameter 2.3\,m filled with gadolinium-loaded liquid scintillator.
  The gadolinium allows for efficient and nearly background-free detection of
  neutrons because of its large thermal cross section and the 8\,MeV of gammas
  released from each capture, which rise well above the background from 
  natural radioactivity.

  \item The \emph{\gcns} is a cylindrical layer with thickness
  0.6\,m filled with liquid scintillator without gadolinium.  Its purpose is to
  provide an interaction region for gammas produced in the \nt.

  \item The \emph{\bufns} is a 1.1\,m layer of non-scintillating 
  mineral oil surrounding the \nt and \gc.
  In it are mounted 390 10" PMTs.  The oil shields
  the scintillator from PMT glass radioactivity as well as externally
  produced gammas and neutrons.

\lu

The \nt is separated from the \gc by a clear acrylic vessel with thickness of
8\,mm, and the \gc from the \buf by an acrylic vessel of thickness 12\,mm.
The \nt, \gc and \buf are collectively called the \emph{\idns}.

The \id is contained in a steel vessel, outside of which is a 0.5\,m thick
layer of liquid scintillator called the \emph{\ivns}, in which 78 8" PMTs are
mounted.  The inner veto is, in turn, surrounded by 15\,cm of passive steel
shielding.  Above the steel shield, the \emph{\ovns}, consisting of several
large-area plastic scintillator panels, provides additional muon tagging and
tracking. 

The far detector is located under a hill with 300 meters water equivalent
overburden. The muon rate through the \iv is 40\,Hz, of which about 0.3\% stop
in the \nt or \gc.  Only data from the far detector is included in this
analysis.  The far detector has been running since 2011.  The first 489.5 days
of livetime are used in this analysis.

Muon capture can be observed with, in principle, up to a seven-fold
coincidence:

\ol

\item \label{sig:track} Muon track

\item \label{sig:xray} Muonic x-rays and Auger electrons from the atomic cascade:
$\ll 1\,$ns after stop

\item \label{sig:recoil} Capture nuclear recoil: O(1$\,\mu$s) after the atomic
cascade

\item \label{sig:deex} $\gamma$/charged particles from nuclear de-ex\-ci\-ta\-tion:
\emph{usually} $\ll1\,$ns after capture, but $>1\,\mu$s in some cases

\item \label{sig:n} Capture of neutrons from nuclear de-excitation:
10--100\,$\mu$s after $\mu$ capture

\item \label{sig:beta} $\beta$ or \betan decay of daughter nucleus:
10\,ms--10\,s after $\mu$ capture

\item \label{sig:betan} Capture of neutron from \betan decay: 10--100\,$\mu$s
after \betan decay

\lo

Signals \ref{sig:track}, \ref{sig:n}, \ref{sig:beta} and \ref{sig:betan}
are easily observable in Double Chooz. In particular, neutrons (5 and 7) can be
detected by the 2.22\,MeV gamma resulting from capture on hydrogen in either
the \nt or \gc, or by the 8\,MeV of gammas resulting from capture on gadolinium
in the \nt. The efficiency for observing these neutrons is acceptable, but
lower than the neutron efficiency of the Double Chooz neutrino analysis because
the trigger is not fully efficient following large signals such as muons.

Signal \ref{sig:deex} with gammas is observable with some complications due to
muon-induced deadtime and baseline shifts.  Signal \ref{sig:xray} is not
observable because it is always prompt and completely masked by the muon
track's light.  It would be easily observable in a segmented detector, however,
at least for high Z targets (for instance, gadolinium gives muonic x-rays at
4.7\,MeV and 1.8\,MeV, and about 1\,MeV of softer x-rays and Auger electrons).
Signals \ref{sig:recoil} and \ref{sig:deex} with charged particles are not
visible because the scintillation is highly quenched.

In Double Chooz, therefore, the coincidence can be up to 5-fold, but as it
turns out, no reactions using all 5 signals are practical to observe.  Examples
do exist for 1+6 (track+$\beta$, such as \is{12}B produced by \is{12}C,
\sect{sec:b12analysis}), 1+5+6 (\dots with a neutron, such as \is{12}B produced
by \is{13}C, same section), 1+6+7 (track+\betan, such as \is{9}Li produced by
\is{13}C, see \sect{sec:betan}), 1+5+6+7 (\dots with a neutron, such as \is{17}N
produced by \is{18}O, same section), or 1+4+6 (track+gamma+$\beta$, excited
states of \is{12}B, see \sect{sec:b12gamma}).

\subsection{Candidate Isotopes}

The most common isotope in Double Chooz is \is{12}C, but \is{13}C, \is{14}N,
\is{16}O, \is{17}O, and \is{18}O are also present in relevant quantities.
While there is also hydrogen in the detector, \mum never undergoes nuclear
captures on it.  In the case that the muon captures atomically on hydrogen, it
is always quickly picked off by a heavier element~\cite{measday}.

Some previous data exists for partial capture rates in \is{12}C, \is{14}N and
\is{16}O, as summarized in \rf{measday}. Some newer results on nitrogen and
oxygen are also available \cite{measdayn,measdayconf}.  For \is{13}C, no
previous measurements of partial capture rates are available.

Tables~\ref{tab:iso} and~\ref{tab:isomore} display all isotopes that could
possibly be observed in Double Chooz from muon capture on any stable isotope of
carbon, nitrogen, or oxygen.  This is the same as listing all possible
muon-capture products of \is{18}O, i.e., anything that can be made by removing
zero or more nucleons from \is{18}N.  In \tab{tab:iso}, isotopes analyzed
in this paper are shown non-italicized.  Stable isotopes are shown as long
dashes (\stab{\textrm{a}}{Z}) and those that decay too slowly as dots
\mbox{(\slow{\textrm{a}}{Z})}.  Isotopes that undergo strong decays are not
shown. Isotopes that can only be formed from \is{17}O or \is{18}O are
italicized.  Because of the rarity of these two parent isotopes, their
daughters are not measured, but some are taken into account when measuring
other isotopes.

\isotable
\isotablemore

\subsection{About Carbon}\label{sec:aboutc13}

The \nt and \gc scintillators and acrylic vessels contain 24.0\,t of carbon.
From EA-IRMS measurements, the molar \is{13}C fractions of the Double Chooz
scintillators are known to be 1.0925(1)\% for the \nt and 1.0906(1)\% for the
\gc.  The atomic capture rates for different isotopes of an element are
identical.  The nuclear capture rate on \is{13}C is a few percent lower than
that of \is{12}C~\cite{measday}.  Because the individual relative errors are
0.01\%, these values can be treated as exact for our purposes.  Even the
difference in \is{13}C fractions in the two scintillators is only 0.17\%
relative.  The most precise measurement reported in this paper has a
statistical error over 1\% relative, so the \nt/\gc difference can be neglected
and a mass-weighted average value of 1.0919\% used instead, again treated as
exact.  Overall, it is expected that $(1.01\pm0.06)$\% of the nuclear captures
on carbon are on \is{13}C.

\subsection{About Oxygen and Nitrogen}\label{sec:abouton}

Both the \gc and \nt scintillators contain nitrogen and oxygen due to their PPO
($\mathrm{C}_{15}\mathrm{H}_{11}\mathrm{NO}$) component.  The \nt scintillator
additionally has oxygen in the gadolinium complex
($\mathrm{GdO}_6\mathrm{C}_{33}\mathrm{H}_{57}$) and in tetrahydrofuran
($\mathrm{C}_{4}\mathrm{H}_{8}\mathrm{O}$)~\cite{scint,largescale}.  The
acrylic ($\mathrm{C}_5\mathrm{H}_8\mathrm{O}_2$) vessels also contribute
oxygen. 

The vast majority of the oxygen is in the acrylic vessels.  The \nt vessel,
including supports and stiffeners, contributes 138\,kg of oxygen, while the \gc
vessel contributes 260\,kg.  Since the acrylic is thin, most decays in it
should be observable with only a minor loss in visible energy, with the
exception of those in the \gc vessel that escape into the \buf.  The \nt
scintillator has an additional 21\,kg of oxygen, and the \gc scintillator has
3\,kg.

After deriving the rate of nuclear captures on carbon in
\sect{sec:b12analysis}, these figures will be used to derive an estimate for
the number of nuclear captures on oxygen.  One major difficulty arises.  While
the nuclear capture probability given a \mum bound to an oxygen nucleus is
well-measured, the rate of atomic capture is uncertain. As discussed in
\cite{measday}, there is no good rule for what happens in mixed targets.  A
large systematic error is taken for this uncertainty in \sect{sec:ratesloose}.

Nitrogen is only present in the scintillators, with 4.8\,kg in the \nt and
2.8\,kg in the \gc.  A similar difficulty with regard to the atomic capture
probability exists with nitrogen as with oxygen.  Given the small total amount
of nitrogen, only \is{14}N is important, the contribution of \is{15}N being
completely negligible.

Muons may also capture on gadolinium in the \nt scintillator, or the various
components of steel in the calibration system and phototubes.  However, despite
the higher probabilities of nuclear capture when \mum become bound to these
atoms, 96.3\% in the case of gadolinium~\cite{Suzuki}, the total amounts
available are sufficiently small that these elements can be ignored both as
possible opportunities for measurement and as sources of background for the
measurements presented here.

\section{Rate of Stopping \mum}\label{sec:rateofstoppingmu}

The analyses below determine the number of decay events for various isotopes.
To translate these counts to probabilities, one needs to know the number of
nuclear \mum captures.  This number is obtained by counting reconstructed
stopping muons and using a Monte Carlo simulation to get the \mum/\mup ratio,
giving the number of stopping \mum.  Using previous experimental data, this can
be translated into the number of nuclear \mum captures.

Double Chooz has no capability to distinguish between \mup and \mum tracks.
From studies using the MUSIC~\cite{music} Monte Carlo simulation along with
charge ratio data from MINOS~\cite{minoschargeratio,minoschargerationear} and
L3+C~\cite{l3c}, it is estimated that $(44.1\pm0.3)$\% of stopping muons are
\mum, which gives a 0.7\% normalization error for all measurements herein.
Results from MUSIC were cross-checked using a custom fast Monte Carlo
simulation incorporating the Gaisser model~\cite{Gaisser2012801}, as modified
by Guan et al~\cite{anothercosmicparam}. The error is dominated by the
systematic error from L3+C, with only a small contribution resulting from the
modeling of our overburden.

Muons that stop in the \id scintillator can be identified by selecting events
with a large energy deposition in the \id and energy in the \iv consistent with
a single muon crossing.  By reconstructing candidates under both stopping and
through-going hypotheses and selecting based on the relative goodness-of-fit,
an 80\% pure sample of muons with stopping points in the \nt or \gc can be
obtained. The remaining 20\% are primarily muons that stop in the \buf and
through-going muons that pass through inefficient regions of the \iv. Because
of Double Chooz's underground location, the level of contamination by stopping
pions is negligible. The muon reconstruction algorithm has been described in
detail in Ref~\cite{fidonim}.  The position resolution for the stopping point
is typically 150\,mm in each of $x$, $y$, and $z$.

In order to select a high-purity sample of stopping muons, a series of further
selection cuts is used.  In particular, candidate muons must have reconstructed
stopping points far from the bottom and sides of the \gc since many muons
reconstructed to stop in these regions are in fact through-going.  The stopping
position must have $z > -1.175\,$m and $r < 1.05\,$m, where $(0,0,0)$ is the
center of the \nt. Further, the energy deposition in the \iv must be within a
tight range consistent with the reconstructed crossing, and the reduced
$\chi^2$ of the muon track fit must be less than 2.0.  Finally, stopping muon
candidates are rejected if another muon event precedes them by less than
0.5\,ms.  This ensures that any neutrons produced can be assigned to the
correct muon.

It was found that $(99.72\pm0.19)\%$ of muons passing these cuts were stopping
muons.  This was determined by testing the uniformity of the probability of a
Michel decay following a stopping muon candidate across the selected spatial
region.  It was cross-checked, at lower precision, by verifying that the total
number of observed Michel decays over 10\,MeV was consistent with the selected
stopping muon count. Efficiencies for this study were evaluated using the
Michel spectra for \mum bound to carbon~\cite{michelspectrum} and free
\mup~\cite{michelrad}.  A second cross-check verified the spatial uniformity of
the probability of observing a \is{12}B decay after a stopping muon candidate. 

To test the rate of reconstruction failures directly, a sample of through-going
muon events from our data set was modified to remove \iv hits at the endpoint.
This causes them to resemble muons that stop in the \buf, allowing us to test
the rate at which such muons reconstruct with end points far from the edge.  No
such reconstructions were found, and a limit was set that less than 0.03\% of
the selected sample consists of such misreconstructions.  Because our technique
of hit removal is only an approximation for the real background, this study is
treated as a cross-check for numerical purposes, but gives us confidence that
the purity quoted above is conservative.

In this high purity sample, there are 1\,628\,874 reconstructed stopping muons.
Of the \mum, $(99.8\pm0.1)\%$ become bound in carbon atoms. The uncertainty on
this number derives from the uncertainty on the probability of a muon becoming
attached to non-carbon atoms, as described in \sect{sec:abouton}. From the
purity and \mum fractions above, this gives $(7.149\pm0.054)\times10^5$ \mum
that are available to undergo nuclear capture on carbon.  Given total nuclear
capture probabilities of $(7.69\pm0.09)$\% and $(7.3\pm0.4)$\% for \is{12}C and
\is{13}C~\cite{measday}, respectively, and 489.5 days of livetime, the nuclear
capture rates in the high purity sample are $111.1\pm0.6$/day for \is{12}C and
$1.16\pm0.06$/day for \is{13}C.

\section{\is{12}B/\is{13}B Analysis}\label{sec:b12analysis}

Previous experiments have measured \is{12}B production by stopping muons in
carbon, although none have done so using a method that can separate production
by \is{12}C and \is{13}C~\cite{maier,measday}. Given its short half-life and
high $Q_\beta$ value of (see \tab{tab:isomore}), \is{12}B is readily
observable. This applies to \is{13}B as well, but is quite difficult to
distinguish from \is{12}B given its very similar half-life and beta spectrum.

\subsection{Selection}\label{sec:b12selection}

Beta decay candidates must occur at least 1\,ms after a stopping muon without a
Michel electron.  The efficiency for observing Michel decays is about 78\%, as
determined by comparing the number found in the high-purity sample of
stopping muons to the number expected given the muon charge ratio and inclusive
capture probability.  The stopping muon must be followed by at least 100\,s of
continuous data-taking. 

An event is selected as a beta decay if it is not light noise (spontaneous
emission of light from PMTs, described in~\cite{3rdpub}) and is between 4 and
15\,MeV.  The energy cut has an efficiency of $(85.0\pm0.7)\%$, as found from a
low-background subset of \is{12}B events.  In evaluating this efficiency, the
portion of the \is{12}B spectrum below 0.5\,MeV was filled with a beta
spectrum~\cite{energylevels1990,energylevels1980,b12_1957,b12_1978,Schenter_Vogel_1983,forbiddenbeta,betashapes}
Monte Carlo simulation, taking into account the detector response.  In
addition, it must not occur within 0.5\,ms after any subsequent muon, to avoid
selecting neutron captures as beta decays. 

The overall selection efficiency, including inefficiency due to light noise
removal, the \is{12}B energy cut, Michel electron selection, subsequent muon
veto and the requirement of continuous data-taking, is $(81.0\pm0.6)\%$.  No
position cut is made for the beta decay in order to avoid incurring additional error due
to the efficiency of such a cut.  While this is advantageous for the
measurement of \is{12}B, a position cut will be used in the following sections
for all other isotope searches to reduce background.

This sample of events is divided into those in which the muon is followed by
zero, one, or two neutron captures.  In the zero neutron case, the signal is
either \nrn{\ism{12}C}{}{\ism{12}B} or \nrn{\ism{13}C}{}{\ism{13}B}.  In the one
neutron case, the signal is \nrn{\ism{13}C}{n}{\ism{12}B}.  The two neutron
case exists to account for this reaction when an additional accidental
neutron-like event is selected, which has a probability of $1.1\times10^{-4}$.
The neutron efficiency during the data taking period analyzed here is a
function of the preceding muon energy and has a value, averaged over the \nt
and \gc, of $(57\pm1)\%$.

\borontab
  
An extended unbinned maximum likelihood~\cite{exlike} fit using the event
timing is done simultaneously for these three reactions. The fit takes into
account neutron efficiencies on an event-by-event basis, with nuisance
parameters for production of \is{8}Li (half-life $(839.9\pm0.9)$\,ms),
\is{8}Li+n, \is{9}Li (half-life $(178.3\pm0.4)$\,ms), \is{9}Li+n, \is{16}N
(half-life $(7.13\pm0.02)$\,s) and accidental coincidences.  The lifetimes of
each isotope are allowed to float with pull terms constraining them within
published errors.  The total probability of isotope production from \is{13}C is
constrained to be at most 100\% with a pull term taking into account the error
on the total nuclear capture rate on \is{13}C.  The neutron efficiency is included as a
fit parameter and allowed to float within its errors.

Selected events and the fits are shown in \fig{fig:b12fromc13}.  The results of
the fit for each reaction follow.  For each, three numbers are given: (1) the
probability per atomic capture, which has the least dependence on external
measurements, (2) the capture rate, which relies on previous measurements of
the bound \mum lifetime, $(2028\pm2)$\,ns for \is{12}C and $(2037\pm8)$\,ns for
\is{13}C~\cite{measday}, and (3) the probability per nuclear capture, which
depends on the total nuclear capture probability.  The total nuclear capture
probability has been determined using the bound \mum lifetime, so the external
uncertainties on the capture rate and nuclear capture probability have the same
origin. However, since determining the nuclear capture probability involves
subtracting two similar numbers, $P = (\tau_\mathrm{free} -
\tau_\mathrm{bound})/\tau_\mathrm{free}$, the fractional uncertainty is much
larger.

Results are given in \tab{tab:boron}.  All sources of errors handled in the fit, 
such as the neutron efficiency and external lifetime measurements, are
considered statistical.  Note the asymmetric errors for
\nrn{\ism{12}C}{}{\ism{12}B}. These are due to the possibility of a large
contribution from \is{13}B. Since the best fit for \is{13}B is zero, this
source of error is one-sided. 

The most recent previous measurement of any of these processes comes from
\rf{maier}, in which the probability per atomic capture for
\is{12}C$\rightarrow$\is{12}B is given as $(1.55\pm0.06)\%$.
This was converted by \rf{roesch} to a rate of $(7.05\pm0.27)\times 10^{3}\,\mathrm s^{-1}$ and
subsequently quoted by \rf{measday} as a probability per nuclear capture of
$(18.6\pm0.7)$\%.  Two corrections to these figure are needed.  From the beginning, the
contributions from \is{13}C are neglected, which requires a downward correction.
Second, the rate results from an erroneous division of
the atomic capture probability by the free muon lifetime, rather than the bound
lifetime.  This requires an upward correction.
Our best estimates of the correct figures for comparison are
$(1.51\pm0.07)\%$ for the probability per atomic capture,
$(19.6\pm0.8)$\% per nuclear capture,
and $(7.44\pm0.35)\times 10^{3}\,\mathrm s^{-1}$ for the rate. In all cases, 
our figures are about 2.5$\sigma$ lower.

\btwelvefromcthirteenfig

\section{Total Nuclear Capture Rates}\label{sec:ratesloose}

For all remaining reactions, a looser selection is used for stopping muons to
increase the statistics for processes less common than \is{12}B production.
This selection relaxes the above-described requirements on stopping position,
track fit quality and \iv energy.  Stopping muon candidates are accepted if the
stopping point is not within 35\,mm of the bottom or sides of the \gc and if
the fit reduced $\chi^2$ is less than 10.  The constraints on acceptable \iv
energy are also loosened.

To determine the number of true stopping \mum in this expanded sample, the
ratio of \is{12}B-like events found in the two samples is used.  The
high-purity sample has $7882\pm95$, while $17\,580\pm150$ additional events are
found in the expanded sample.  This gives $(2.309\pm0.032)\times10^6$ atomic
captures on carbon in the new sample.  Some impurity and a larger normalization
error are therefore traded for 3.2 times the statistics.

The two samples have very similar \is{12}C/\is{13}C ratios, as described in
\sect{sec:aboutc13}, however the new sample contains substantially more oxygen
by including the \nt vessel and supports.  There is therefore a concern that
the simple ratio above may be incorrect due to presence of \is{12}B from the
reaction \nrn{\ism{16}O}{\alpha}{\ism{12}B}.  Data on this reaction appears to
be limited to a single emulsion experiment~\cite{chicago1953}, but even under
moderately pessimistic assumptions, this is enough to argue that it has a
negligible effect.

The following rates for the expanded sample are found using the total number of
atomic captures for carbon above and the known total nuclear capture
probability:

\ul
\item \is{12}C
  \ul
  \item $4664\pm65$ atomic captures/day
  \item $358.8\pm6.5$ nuclear captures/day
  \lu
\item \is{13}C
  \ul
  \item $51.5\pm0.71$ atomic captures/day
  \item $3.75\pm0.19$ nuclear captures/day
  \lu
\lu
The error on the \is{12}C nuclear capture rate is shared equally between the
\is{12}B-like event statistics discussed above and the probability of nuclear
capture following atomic capture.  The error on \is{13}C results mainly from
the probability of nuclear capture.  

Rates for other isotopes are estimated using the chemical composition of the
detector.  These are given as ranges to emphasize the large uncertainties
surrounding probability of atomic capture in chemical compounds.  Based on data
presented in \rf{measday}, it is conservatively assumed that oxygen (nitrogen)
overcaptures by a factor of \mbox{1--2} \mbox{(1--1.5)} as compared to its
number density.  The probability within these ranges will be treated as
uniform:

\ul

\item 0.2--0.3 \is{14}N nuclear captures/day

\item 6--13 \is{16}O nuclear captures/day

\item 1--2 \is{17}O nuclear captures for the entire data set

\item 5--12 \is{18}O nuclear captures for the entire data set.

\lu

\section{$\beta$\MakeLowercase{n} Isotopes}\label{sec:betan}

The Double Chooz detector is optimized for the detection of
$\bar\nu_\mathrm{e}$ via the inverse beta decay reaction,
p$(\bar\nu_\mathrm{e},e^+)$n.  The resulting positron gives a prompt signal and
the subsequent capture of the neutron on either gadolinium or hydrogen gives a
delayed coincidence.  A major background to this process is production of
\is{9}Li by cosmic muons, as its \betan decay produces a signal that is nearly
indistinguishable from the inverse beta decay signal.  For this analysis, the
existing inverse beta decay selection is used to select \betan decays in the
\nt and \gc, both for events with captures on gadolinium~\cite{3rdpub} and
hydrogen~\cite{hn}.

\btwelveresfig

\lininedposfig

An excess over background of \betan events is found near stopping muons both in
time and space.  Events for which the reconstructed distance from the muon
stopping point to the $\beta$ decay is less than 300\,mm are selected for the
fit described below.  The efficiency of this cut is estimated using the sample
of \is{12}B events (see \sect{sec:b12selection}) and is found to be
$(92.0\pm0.5)$\% in the \nt and $(74.9\pm0.5)$\% in the \gc (see
\fig{fig:b12res}).  Fifteen \betan-like events are selected within 300\,mm and
400\,ms of stopping muons. The spatial distribution for these events is shown
in \fig{fig:li9dpos}.  The close correlation in space demonstrates that the
isotopes are being produced at the stopping point, rather than along the muon
track. This also excludes the possibility that they are created by correlated
through-going muons.

\lininefig

The time distributions are shown in \fig{fig:li9} for the case in which no
requirement is put on neutron detection following the muon and the case of
selecting one neutron following the muon.  The flat background is formed from
accidental coincidences between stopping muon candidates and \betan-like
events.  It is due almost entirely to reactor neutrino interactions.  As at
least one reactor of the Chooz power plant was active for 98\% of this data
set's livetime, this background is unavoidable.  The efficiency of this
selection for \is{9}Li \betan decays in the \nt with the neutron capturing on
gadolinium is $(82.0\pm0.5)\%$ and for decays in the \gc with the neutron
capturing on hydrogen is $(62.7\pm0.4)\%$.

In the case of no neutron requirement, the time distribution is fit using the
unbinned maximum likelihood method~\cite{exlike} under the hypothesis that the
excess over a flat background consists of a combination of several \betan
isotopes (half-lives given in parentheses): \is{11}Li (8.75\,ms), \is{13}B
(17.33\,ms), \is{8}He (119.1\,ms), \is{9}Li (178.3\,ms), \is{16}C (747\,ms),
and \is{17}N (4.183\,s).  The fit is done in three bins of reactor power to
take advantage of the lower background rates during zero- and one-reactor
periods.  Some potentially-present \betan isotopes were excluded from the fit.
First, \is{14}Be, \is{15}B and \is{12}Be were excluded because their half-lives of
4.35\,ms, 9.93\,ms and 22\,ms are close to those of \is{11}Li and \is{13}B, and, as
will be shown in \sect{sec:li11}, there is no evidence of an O(10\,ms)
component.  Second, \is{17}C was excluded because its half-life of 193\,ms is
very close to \is{9}Li's and it can only be produced by
\is{18}O$(\mu^-,\nu\mathrm p)$\is{17}C. With a \betan branching ratio of 32\%,
it is expected to contribute much less than 1 event for any reasonable estimate
of the $(\mu^-,\nu\mathrm p)$ probability.  Third, \is{18}N is excluded because
its half-life of 620\,ms is very close to \is{16}C's and with a \betan
branching ratio of only 7\%, the expected number of events is less than one
regardless of the probability of \nrn{\ism{18}O}{}{\ism{18}N}.

With all six isotopes free in the fit, there is a 5.5$\sigma$ preference for a
\betan signal over the background-only hypothesis.  The fit is unable to
constrain the probabilities of producing the isotopes with half-lives longer
than \is{9}Li's due to the amount of accidental background.  With all isotopes
free to take any value in the fit, significant degeneracy exists between these
and the shorter lived isotopes.  In order to be able to measure the \is{9}Li
and \is{8}He probabilities, the contributions from \is{16}C and \is{17}N are
therefore given pull terms constraining their probabilities to $(5\pm5)$\% and
$(50\pm25)$\%, respectively, based on previous measurements of similar
reactions in other isotopes~\cite{measdayn,measday}.  This done, the
significance of \is{9}Li and \is{8}He specifically being present is
\linineheeightsignificance.  As shown in \fig{fig:li9cont}, there is little
power to distinguish these two isotopes from each other.  

As shown in \fig{fig:li9energy} the beta spectrum of selected \betan events is
compatible with that of \is{9}Li plus the reactor neutrino background.  While
the statistics do not allow a detailed fit, the event at 9.47\,MeV provides
further evidence against the hypothesis that neutrinos or other backgrounds are
the only events present, since only a very small fraction of reactor neutrinos
reach this energy, and the rate of other neutrino backgrounds is small compared
to the neutrino rate.  It is also incompatible with all events being \is{8}He,
\is{16}C, \is{17}C, or \is{17}N (see \betan endpoint in \tab{tab:iso}).

Assuming no \is{8}He, the probability of forming \is{9}Li from a muon capture
in \is{\mathrm{nat}}C is found to be
$(2.4\pm0.9\mathrm{(stat)}\pm0.04\mathrm{(syst)})\times10^{-4}$.  The
systematic error is dominated by the uncertainty on the number of \mum\
captures.  If no constraint is put on \is{8}He, only limits can be set on it
and \is{9}Li. The \is{9}Li limit with \is{8}He unconstrained is
$<4\times10^{-4}$ at 90\% CL.  For \is{8}He it is $<7\times10^{-4}$ at 90\% CL.

If \is{9}Li were produced via \nrn{\ism{12}C}{\mathrm{n2p}}{\ism{9}Li}, most
events would have an observable neutron capture between the stopping muon and
the $\beta$ decay.  Because only a minority of stopping muon candidates satisfy
this demand, requiring observation of this neutron reduces the neutrino background considerably.  Efficiency for
observing these neutrons is reduced by deadtime following muons during the data
period used in this paper; it is 57\% for the \nt and 87\% for the \gc. 

In fact, few of the events survive the neutron requirement, as shown in
\fig{fig:li9}.  The remaining events give an overall significance of \betan
production of 3.4$\sigma$ and are compatible with the expected rate of
\nrn{\ism{18}O}{n}{\ism{17}N}.  The fitted amount of \is{9}Li+\is{8}He in the
two samples is incompatible with the hypothesis of
\nrn{\ism{12}C}{\mathrm{n2p}}{\ism{9}Li} at the 2.7$\sigma$ level, indicating
that the reaction \nrn{\ism{13}C}{\alpha}{\ism{9}Li} most likely is responsible
for the \is{9}Li produced by stopping muons.  Under this interpretation, and
assuming no \is{8}He production, the probability for production from \is{13}C
is $(2.4\pm0.9\mathrm{(stat)}\pm0.05\mathrm{(syst)})\%$.

\lininecontfig

\subsection{\is{11}Li and \is{13}B}\label{sec:li11}

The above analysis also constrains \is{11}Li and \is{13}B, which would appear
as early \betan events given their half lives of 8.75\,ms and 17.33\,ms,
respectively. Both would be produced from \is{13}C without neutrons.  The
constraint on \is{13}B is independent of that found in \sect{sec:b12analysis},
since in that fit the $\beta$ mode was used.

The efficiency for selecting \is{11}Li is reduced to 92.7\% compared to
\is{9}Li due to events lost in the 1\,ms muon veto.  The fraction of \is{11}Li
that undergoes \betan decay is 83\%. The efficiency for selecting \is{13}B is
likewise 96.1\% relative to \is{9}Li. Its \betan branching fraction is only
0.29\%. 

The procedure described in the previous section is repeated, but requiring no
neutrons.  The fit finds no evidence for either isotope (see
\fig{fig:b13li11}). Because of the low probability of \is{13}B-\betan, this fit
adds no significant information to the result of the \is{12}B/\is{13}B fit
described in previous sections. The \is{11}Li probability, allowing the
\is{13}B contribution to float freely, is $<0.7\%$ at 90\% CL.

\linineenergyfig
\bthirteenlielevenfig

\section{Decay Chain: \is{12}B\MakeLowercase{e} }\label{sec:decaychain}

The isotope \is{12}Be is a special case because it decays via a 2-step chain to
\is{12}C via \is{12}B.  The two beta decays are very similar in energy and
lifetime.  The first decay (\is{12}Be$\rightarrow$\is{12}B, half-life
21.49\,ms, $\beta$ endpoint 11.7\,MeV) is selected with a similar procedure as
used to find \is{12}B in \sect{sec:b12analysis}.  The decay is required to
occur 1--100\,ms after the stopping muon (92.9\% efficiency, using the
\is{12}Be lifetime), within 400\,mm of its stopping point (91.6\% efficiency,
as judged by the sample of \is{12}B decays, as in \sect{sec:betan}), and have
energy 3--12\,MeV ($89\pm2$\% efficiency, based on Monte Carlo simulation).
The overall selection efficiency for the \is{12}Be decay is $(75\pm2)$\%.

If a \is{12}Be candidate is found, the selection is then repeated to search for
the \is{12}B decay.  The distance cut is applied relative to the \is{12}Be
decay candidate, since the position reconstruction for point-like events has
higher resolution than that for muons.  The \is{12}B decay is required to occur
1--150\,ms after the \is{12}Be decay (96.0\% efficiency), within 400\,mm (97\%
efficiency), and have energy 3--15\,MeV ($92.5\pm0.7$\% efficiency).  The
\is{12}Be and \is{12}B decays are not allowed to occur within 1\,ms of each
other to avoid selection of inverse beta decay events or \is{12}B produced by
neutron spallation.  The overall selection efficiency for the \is{12}B is
$(85\pm1$)\%.  The selection efficiency for the decay chain is $(63\pm2$)\%. 

No events are selected.  An upper limit is set of \mbox{$<0.20\%$} per \is{13}C
capture at 90\% CL.

\section{$\beta$ Isotopes}\label{sec:beta}

\subsection{\is{8}L\MakeLowercase{i}}\label{sec:li8}

For \is{8}Li, for which the signal is a single beta decay (half-life 839.9\,ms,
$\beta$ endpoint $\sim$13\,MeV), the same analysis as was used to measure
\is{12}B in \sect{sec:b12analysis} is used, but with modifications to reduce
background.  The \is{8}Li decay is required to have an energy of 5--14\,MeV and
be within 400\,mm of the stopping muon.  Events are selected with and without
neutrons following the muon.  Events with a neutron are interpreted as
\nrn{\ism{13}C}{n\alpha}{\ism{8}Li} and those without as
\nrn{\ism{12}C}{\alpha}{\ism{8}Li}.

Since the \is{8}Li lifetime is fairly close to the \is{9}Li lifetime, inverse
beta decay candidates are excluded from the sample to reduce contamination from
\is{9}Li.  The remaining 49.2\% of \is{9}Li that undergo plain $\beta$ decay are not
affected by this cut.  The small amount of remaining \is{9}Li is constrained
with pull terms derived from the \betan analysis.  

To suppress background caused by muon-induced \is{12}C$(\mathrm{n,p})$\is{12}B,
a likelihood classifier is used that takes as input the distance between a
candidate decay and the most recent through-going muon and the number of
neutron captures following that muon~\cite{3rdpub}. This classifier removes 50\% of the
accidental background and has a signal efficiency of 90.6\%.

The selected events and fit results are shown in \fig{fig:li8}.  If the
\is{8}Li half-life is allowed to float without a pull term in the zero (one)
neutron sample alone, the result is $870^{+120}_{-100}$\,ms
($750^{+260}_{-270}$\,ms), as compared to the accepted value of 839.9\,ms. 

\lieightfig

The probability of \nrn{\ism{12}C}{\alpha}{\ism{8}Li} per nuclear capture is
found to be \probEightLiFromTwelveC, where the systematic error is dominated by
the uncertainty on the \is{8}Li energy cut efficiency.  The probability of
\nrn{\ism{13}C}{\alpha n}{\ism{8}Li} per nuclear capture is found to be
\probEightLiFromThirteenC, where the systematic error is dominated by the
uncertainty on the fraction of muons that undergo nuclear capture on \is{13}C.
It should be noted that since Double Chooz is nearly blind to heavy charged
particles, these probabilities are not correct if 
reactions such as \nrn{\ism{12}C}{n\ism{3}He}{\ism{8}Li} or
\nrn{\ism{12}C}{ndp}{\ism{8}Li} occur at any significant rate.

\subsection{Isotopes Produced with Many Neutrons}\label{sec:manyneutrons}

The isotopes \is{8}B, \is{12}N and \is{9}C must be produced via the emission of
4 or more neutrons by the parent nucleus following muon capture, which allows
for a nearly background-free search.   First, \nrn{\ism{12}C}{4n}{\ism{8}B} is
considered.  The resulting \is{8}B has a half-life of 770\,ms and a \betap endpoint of
$\sim$14\,MeV (it decays to a broad resonance of \is{8}Be), providing up to
15\,MeV of visible energy. The average efficiency for observing all four
neutrons is $(48\pm2)$\%.  Candidate \is{8}B events are selected with
4--18\,MeV, which has $(97\pm1)$\% efficiency due to the positron annihilation
following the \betap decay.  Candidates are required to be within 400\,mm of
the stopping muon.  These energy and position cuts are shared with the other
isotopes in this section.  The overall efficiency is $(40\pm2)$\%.  No events
are selected, and a limit of $<3.2\times10^{-5}$ per nuclear capture on
\is{12}C at 90\% CL is set.

Next, \nrn{\ism{16}O}{4n}{\ism{12}N} is examined.  With a \betap end-point of
17.3\,MeV and half-life of only 11\,ms, we restrict the search to 1--100\,ms
and can allow events with either 3 or 4 observed neutrons without admitting
significant background. As \is{12}N would be produced primarily in the acrylic,
and the neutron efficiency is a function of position, this efficiency is reevaluated and
found to be 70\% for seeing at least 3 out of 4 neutrons.  The overall
selection efficiency is 54\%.  No events are selected.  A limit is set of
$<8\times10^{-4}$ per capture on \is{16}O at 90\% CL.  The systematic error on
the number of oxygen captures is taken into account in this limit using the
method described in \rf{cdf5928} taking the prior to be flat in the range
quoted in \sect{sec:ratesloose}.

Finally, we search for \is{9}C. Possible production mechanisms include
\nrn{\ism{16}O}{p6n}{\ism{9}C}, \nrn{\ism{16}O}{d5n}{\ism{9}C}, and
\nrn{\ism{14}N}{5n}{\ism{9}C}.  The half-life of \is{9}C is 127\,ms and it has
a \betap endpoint of 15.5\,MeV. Events are selected between 1\,ms and 1\,s if
at least four neutrons are observed after the muon.  In the case of oxygen, it
is conservatively assumed that the true number of neutrons is five.  Given this
assumption, the neutron efficiency is 64\% for an overall efficiency of 53\%.
With no signal and no background, a limit of $<9\times10^{-4}$ at 90\% CL is
set.  In the case of nitrogen, the neutron efficiency is 63\% and overall
efficiency is 55\%.  A limit is set of $<3.6$\% per nuclear capture on \is{14}N
at 90\% CL.

\subsection{Long-lived Isotopes: \is{15}C \& \is{11}Be}\label{sec:n16}

\beelevenfig

The isotopes \is{15}C, \is{16}N and \is{11}Be are those with the longest
half-lives included in this analysis, being 2.449, 7.13 and 13.81\,s,
respectively.  Since the fraction of captures on \is{16}O that produce \is{16}N
is already known to be $(11\pm1)$\%~\cite{measday}, it is treated as a
background for the other isotopes.  It and \is{15}C can be produced by oxygen
in the acrylic, while \is{11}Be is produced from carbon, primarily in the
scintillator.

Accidental background limits our ability to observe long-lived isotopes.  For
this search, a very tight spatial cut is made, restricting the decay candidate
to be within 200\,mm of the muon.  Events are vetoed if any neutrons are
observed.  The candidate must be at least 1\,ms after the most recent muon
anywhere in the detector and at least 0.1\,ms after the previous trigger. The
\is{12}B likelihood (see \sect{sec:li8}) is applied.  The decay energy is
restricted to 4--12\,MeV.  The  efficiencies are 33\% for \is{11}Be, 27\% for
\is{15}C and 27\% for \is{16}N. 

A joint fit for \is{12}B, \is{8}Li, \is{15}C, \is{16}N, and \is{11}Be is
performed.  With all isotopes freely floating, the fit favors the presence of
some combination of \is{15}C, \is{16}N and \is{11}Be, as compared to none, at
6.4$\sigma$ (see \fig{fig:be11}).  The fit prefers to explain nearly all of the
excess as \is{16}N, finding $7.2^{+1.3}_{-3.4}\times10^2$ \is{16}N events,
corrected for efficiencies, compared to an expectation of
(3.0--7.0)$\times10^{2}$ given the amount of oxygen in the detector and the
known \is{16}N production probability. 

To produce the final results for \is{11}Be and \is{15}C, a pull term is used
for \is{16}N, constraining it to the expected number of events within errors.
No significant amount of the other isotopes is found.  Limits are set of
\mbox{$<0.20\%$} for \nrn{\ism{12}C}{p}{\ism{11}Be} at 90\% CL and $<9\%$ for
\nrn{\ism{16}O}{p}{\ism{15}C} at 90\% CL.

\subsection{\is{6}He}\label{sec:he6}

\hesixfig

The half-life of \is{6}He is very close to \is{8}Li's (801 and 840\,ms,
respectively), but it has a much lower $\beta$ endpoint of 3.51\,MeV.  When
fitting for \is{8}Li above, any possible contribution from \is{6}He is removed
by the energy cut.  To search for \is{6}He, however, it is necessary to fit the
spectrum, setting this analysis apart from all others reported on in this
paper. 

To search for a correlation between low energy \is{6}He decays and their parent
muons, a complex fit is needed to reduce the accidental background.  Most of
this background results from gamma decays outside the \gc, such as in the PMTs
and surrounding rock, and so drops dramatically as one moves towards the center
of the \nt.  Therefore, a binned fit is done in energy and position.  There are
five spatial bins, the first four of which have approximately equal volume:

\ol

\item $r < 740$\,mm and $|z| < 740$\,mm (2.546\,m$^3$), where $r$ is horizontal
distance from the center of the \nt and $z$ is the vertical distance.

\item $r < 933$\,mm and $|z|<933$\,mm, excluding the innermost volume
(2.557\,m$^3$)

\item $r<1068$\,mm and $|z|<1068$\,mm, excluding the innermost two volumes
(2.551\,m$^3$)

\item $r > 1068$\,mm or $|z| > 1068$\,mm and inside the \nt (2.759\,m$^3$)

\item In the \gc (22.74\,m$^3$)

\lo

Each spatial bin is binned in energy with 0.125\,MeV bins in the range
0.375--15\,MeV.  The following selection is used: The time since the stopping
muon must be at least 0.3\,s and not more than  1.602\,s (two \is{6}He
half-lives).  The lower bound reduces the background from correlated \is{12}B
to 1.1 events for the whole data set.  The decay must be within 200\,mm of the
muon stopping point, at least 0.1\,ms since the last trigger, at least 1\,ms
since the last muon, and must not have a high likelihood of being \is{12}B
produced by a through-going muon.  In four separate analyses, the muon must be
followed by exactly 0, 1, 2, or 3 neutrons.

\hensixfig

\hesixtab

The \is{6}He efficiency after these cuts ranges from 25\% for the zero-neutron
case to 10\% for the three-neutron case.  The fit is done with 16 free
parameters:

\ul

\item[(1)] The amount of \is{6}He.

\item[(5)] The accidental background normalization in each region.  This
background is measured using an offtime window.  The normalization is
constrained by a pull term to be within the statistical errors of the off-time
sample.

\item[(5)] The amount of \is{8}Li in each region, which is the dominant
background at the \is{6}He endpoint for the inner regions.

\item[(5)] The amount of \is{16}N in each region.  This is allowed to float
freely in each region since its 6.1\,MeV gamma means that the number
reconstructed in each region is not simply the fraction produced there.  The
inclusion of \is{16}N is not very important, but accounts for the 6--7\,MeV
peak seen in the plots and avoids slightly overestimating the \is{8}Li
background in the signal region.

\lu

The results are shown in \figs{fig:he6} and \ref{fig:he6nn} and \tab{tab:he6}.
There are no clear signals, but the 1 neutron case does have an excess.
Without claiming to have discovered this process, we note that
\nrn{\ism{12}C}{p\alpha}{\ism{7}He} would give this signal since \is{7}He
immediately decays via neutron emission to \is{6}He.  In contrast, the 0, 2,
and 3 neutron final states would require less favorable combinations of charged
particle emission.

\subsection{\is{14}B}\label{sec:b14}

With its 12.5\,ms half-life and extremely high Q value of 20.6\,MeV, \is{14}B
can be identified with very little background by looking for early events above
the \is{12}B $\beta$ endpoint.  Events are therefore selected if the candidate
decay has 15--22\,MeV, occurs between 1\,ms and 62.5\,ms after the stopping
muon, and within 400\,mm of the stopping point.   The energy cut efficiency is
evaluated using beta branch probabilities found at TUNL~\cite{tunl} and is
found to be 39\%.  The overall efficiency is 31\%.  No events are observed with
no background (see \fig{fig:b14}), giving a limit of $<0.16\%$ at 90\% CL for
\is{16}O$(\mu^-,\nu 2\mathrm{p})$\is{14}B, including a systematic error on the
number of oxygen captures.

See \tab{tab:summary} for a summary of isotope production results.

\bfourteenfig

\summarytabletable

\section{Exclusive states of \is{12}B}\label{sec:b12gamma}

When \is{12}B is formed in an excited state, it is possible to observe gammas
from its de-excitation.  This allows measurements both of the rate of
transition to excited states and, by subtraction, of the ground state.  Because
Double Chooz has $4\pi$ coverage and good gamma containment, each state can be
identified unambiguously, despite the ``unbelievably
capricious''~\cite{measday} arrangement of bound levels, giving nearly
degenerate individual gamma energies, that has plagued earlier measurements.  

The states of \is{12}B are well known, with three relevant bound levels at
951\,keV, 1674\,keV and 2621\,keV, each of which has previously been shown to
be populated by muon capture on \is{12}C.  As will be shown, our data suggests
that some unbound levels are also populated at a measurable rate, including
3759\,keV.

\subsection{Selection}\label{sec:gammaselection}

In order to only use events for which the trigger has been shown to be fully
efficient and the event energy reliable, only stopping muons depositing less
than 215\,MeV in the \id are used, this being $(38.6\pm0.4)\%$ of stopping
muons.  Also due to the behavior of the trigger, the times of the events are
limited as follows:

\ol

\item When studying 953\,keV and 1674\,keV levels, the capture must be
$4\,\mu\mathrm s \le t < 5\,\mu$s after the muon.

\item For 2621\,keV and 3759\,keV, $3\,\mu\mathrm s \le t < 5\,\mu$s.

\item To search for higher levels: $2\,\mu\mathrm s \le t < 5\,\mu$s.

\lo

Using the bound muon lifetime, these timing requirements have efficiencies of
5.5\%, 14.3\% and 28.6\%, respectively.

To reduce background from other processes, notably
\nrn{\ism{12}C}{n}{\ism{11}B^*}, it is necessary to require observation of the
\is{12}B beta decay as well as the gammas.  These events are selected in a
similar way as in \sect{sec:b12analysis}, with the additional requirements that
the decay be between 2--60\,ms after the stopping muon and reconstruct within
400\,mm of the muon endpoint.  By requiring the beta decay to occur 2\,ms after
the muon stop, we avoid selecting neutron captures as \is{12}B. 

Given these requirements, the overall efficiency is 1.09\% for selection 1,
4.14\% for selection 2 and 6.8\% for selection 3, with 1.5\% relative error on
each, with the requirements on muon energy and gamma timing being the main
drivers.

Selected events are shown in \fig{fig:b12gammafit}.  The 2621\,keV and 952\,keV
levels are clearly visible.  The 1674\,keV level is not so clear, but
fortunately is already known to be present.  

There are also four events in selection~2 between 3500\,keV and 4000\,keV,
which may be due to the 3759\,keV level. This level is known to de-excite via
gamma emission some fraction of the time~\cite{b12levels}. It is therefore
included in the fits below.  Further, in selection~3 there are six events
between 5800\,keV and 7400\,keV, which also rises well above background.  One
possibility is that these originate from the 6600\,keV level of \is{12}B, but
as this assignment seems less clear, we simply note that it appears that other
states are populated to some extent.  Similarly, the five events between
10\,MeV and 25\,MeV are significantly over the background of 0.15 events.

\subsection{Fit}\label{sec:gammafit}

Figure~\ref{fig:b12gammafit} shows the fits used to extract the number of
events at each level.  The fit for selection 1 is used to find the number of
953\,keV and 1674\,keV events, while that for selection 2 is used to find the
number of 2621\,keV and 3759\,keV events.  The fit for selection 3 is used only
to demonstrate the level of consistency of the higher energy data with
expectations.  The fit includes terms for each of the following, which will be
explained in detail in the following sections:

\ul

\item The signal: gammas from
\is{12}{C}$\rightarrow{}^{12}\mathrm{B}^*\rightarrow{}^{12}\mathrm{B_{g.s.}}
\gamma$

\item Accidental background from the sequence (1) stopping muon track (2) muon
decay (3) accidental \is{12}{B}-like event.  Event (2) is selected as the
gamma, giving a Michel spectrum.

\item Accidental background from the sequence (1) stopping muon track --- the
muon creates \is{12}{B} in the ground state (2) accidental gamma-like event (3)
\is{12}{B} decay.  This gives an accidental spectrum mostly at low energy from
detector radioactivity.

\item Correlated background from \is{13}{C}$\rightarrow$\is{8}Li$^*$.

\item Correlated background from \is{13}{C}$\rightarrow$\is{12}{B}$^*$+n, where
the neutron is lost.

\item Correlated background from
\is{13}{C}$\rightarrow$\is{12}{B}$_\mathrm{g.s.}$+n, in which the neutron
captures early and is selected as the gamma.

\item Background from through-going muons producing \is{12}{B} and several
neutrons via spallation.

\lu

Following the sections on each background, additional systematics are
discussed.

\btwelvegammafitfig

\bgammacompfig

\subsubsection{Accidentals}\label{sec:gammaaccidentals}

The accidental spectrum was derived by repeating the selection with the timing
window for \is{12}B beta decays shifted to be a long time after the muon.  Many
windows were used to give a high-statistics determination of the spectrum. The
accidentals characterized in this way include both the case of an uncorrelated
gamma (up to $\sim$3\,MeV) with a real \is{12}{B} decay, and the case of a
Michel electron with an uncorrelated \is{12}{B} candidate.

\subsubsection{Correlated Background: \is{8}{Li}}\label{sec:gammacorrbg}

From \sect{sec:li8}, it is known that \is{8}Li is produced in significant
quantities.  It has one bound level at 981\,keV.  No previous results are
available on the rate of transition to particular states.  Beta decays of
\is{8}Li will be selected with about the same efficiency as \is{12}B if they
fall into the 2--60\,ms selection window.  However, the much longer half-life
of \is{8}Li puts only a small fraction in this window.  Because \is{8}{Li} is
produced with an alpha, the visible energy is the sum of that from the alpha
and gamma.  Given our measurements of scintillator quenching, and assuming a
mean alpha energy of 4.5\,MeV~\cite[figure 5b]{chicago1953}, the total visible
energy expected is 1.23\,MeV.

The contribution from \is{8}Li(981) is determined by selecting events as in
\sect{sec:gammaselection}, but with a beta decay window of 0.3--4.2\,s after
the muon.  The significance of the presence of these gammas is 2$\sigma$.
Limits are obtained, and are shown in \tab{tab:gammaeightsummary}.  The central
value, scaled to the number expected in the \is{12}B time window, is input into
the \is{12}B fit and the errors used as a pull term.

Other isotopes were considered for this treatment.  No significant background
to the three bound \is{12}B levels from \is{13}B is expected, since
\is{13}{B}'s first excited state is at 3483\,keV.  It is possible that the four
events tentatively attributed to \is{12}B(3759) are instead \is{13}B(3483); the
energies are more compatible with the former, but not enough information is
available to confidently distinguish the possibilities.

Other known products of muon capture are disfavored because their long
lifetimes and/or low production rates will reduce their contribution to
negligible levels.  For instance, only 0.6\% of \is{16}{N} beta decays will be
within the selected time window.  The bound levels of \is{16}N's are all below
400\,keV.  It has unbound levels at 3353\,keV, 3523\,keV, and 3963\,keV, but
even if these accounted for 100\% of \is{16}N production, the background
contribution would be well under one event, particularly since they are known
to de-excite primary by neutron emission.

There is one bound level in \is{9}{Li}, at 2691\,keV, but because the
production rate is so low, even if it was produced in this state 100\% of the
time, we would only expect to select about 0.3 events.  Other isotopes yield
even lower background rates regardless of what assumptions are made.

\subsubsection{Correlated Background:
\is{13}C$\rightarrow$\is{12}{B}+n}\label{sec:gammab12n}

No previous results are available for the reaction
\nrn{\ism{13}C}{n}{\ism{12}B}, so there is no constraint on the exclusive rates
to each \is{12}B state in this process.  In the case of this analysis, the
neutron efficiency is high, $(85.8\pm1.0)$\%, and the probability of selecting
an event as a neutron when no true neutron is present is low,
$1.1\times10^{-4}$, which means the \is{12}C and \is{13}C variants can be
largely separated.

Events with and without neutrons are therefore both selected and a simultaneous
fit is performed with both samples, with \is{13}C and \is{12}C processes on
equal footing.  The fit includes the neutron efficiency as a parameter, which
is constrained by a 1\% pull term.  The fit results for
\nrn{\ism{13}C}{n}{\ism{12}B} are shown in \tab{tab:gammathirteensummary}.
Because there are so few events, none of the lines is confidently observed.
However, upper limits can be set.  No unitarity condition is applied (i.e. the
fit allows the levels to add up to more than 100\% of the \is{12}{B}+n
production), making the limits conservative.

Assuming no unaccounted-for background, the total fraction of \is{12}{B}
produced from \is{13}{C} in an excited state is
$(78^{+34}_{-21}\mathrm{(stat)}\pm8\mathrm{(syst)})$\%.  This is much higher
than the probability for \is{12}{B} produced from \is{12}{C} to be in an
excited state, but the production mechanism is different so they cannot be
compared directly.  In a naive shell model, \nrn{\ism{13}C}{n}{\ism{12}B}
involves removing a proton from an inner shell 2/3 of the time, whereas in
\nrn{\ism{12}C}{}{\ism{12}B}, no nucleons are removed.

\gammasummarytable
\gammathirteensummarytable
\gammaeightsummarytable

\subsubsection{Background: Spallation}\label{sec:gammathru}

The following sequence of events can also contribute a background at the \hn
and \gdn energies and would appear in the \nrn{\ism{13}C}{n}{\ism{12}B} sample:
(1) through-going muon, selected as a stopping muon, that produces three or
more neutrons through spallation; one makes \is{12}{B} through the $(\mathrm
n,\mathrm p)$ reaction (2) another neutron captures on hydrogen or gadolinium
and is selected as a gamma (3) a third neutron capture is correctly selected as
such (4) the \is{12}{B} beta decays.

Lacking an estimate for the rate of this sequence, it is taken into account by
including free parameters in the fit that control contribution at the \hn and
\gdn energies in the \is{12}B+n sample.  There does seem to be \gdn present,
primarily visible in selection~3 (with one event in selection~2), but the fit
finds no \hn component in any selection. The uncertainty on this is propagated
through the fit. 

Since this sequence is therefore observed to be rare, and spallation typically
produces many neutrons, the corresponding sequence without event (3), the third
neutron capture, which would instead contaminate the
\nrn{\ism{12}C}{}{\ism{12}B} sample, is assumed to be negligible, particularly
because it is much more likely for a single neutron to be captured during the
longer neutron selection time window than the gamma time window.

\def\linenumberfont{\normalfont\tiny\sffamily}

\subsubsection{Correlated Background: \nrn{\ism{13}C}{n}{\ism{12}B} neutrons
selected as gammas}\label{sec:ncontam}

If the neutron from \nrn{\ism{13}C}{n}{\ism{12}B} captures within the gamma
time window, the event will be selected as \nrn{\ism{12}C}{}{\ism{12}B} with a
gamma energy of 2225\,keV.  Using the measured rate of this reaction from
\sect{sec:b12analysis} and the probabilities of capture within the gamma timing
window, about 1.2 \hn events and 0.7 \gdn events are expected for selection 1,
2.4+1.3 for selection 2, and 3.5+1.7 for selection 3.  These rates are included
in the fit as fixed amounts.  They are not allowed to vary in the fit because
any \is{12}B levels below 10\,MeV that are not included in the fit will
spuriously pull the neutron rate up.  Instead, a systematic error is assigned
after the fact, as shown below.

As a cross-check on the rate of \hn events contaminating the 2621\,keV line,
the timing distribution of these gamma candidates was fit for a combination of
the bound \mum lifetime and the neutron capture lifetime.  No evidence of a
neutron timing component was found, with a limit from timing alone of less than
25\% neutron contamination at 90\% CL.

\subsection{Results and Systematics}\label{sec:gammaresults}

Numerical results are shown in \tabs{tab:gammasummary},
\ref{tab:gammathirteensummary} and \ref{tab:gammaeightsummary}.  These include
the rate for ground state \nrn{\ism{12}C}{}{\ism{12}B}, which is found by
subtracting the excited state rates from the total rate.  To cover the
possibility that the 3759\,keV measurement is spurious, its contribution is
allowed to vary from zero to the stated value in this calculation.  Likewise,
to cover the possibility that higher energy states also cascade down to the
ground state, an error is added that covers the possibility that all the events
from 5--30\,MeV in \fig{fig:b12gammafit} are such events.  Both of these
effects are small compared to the statistical error.

In the tables, all sources of error handled directly in the fit are considered
to be statistical.  The additional systematic errors are as follows:

\ul

\item For results stated as a fraction of \is{12}{B} production, many
systematic errors cancel, leaving an efficiency error of 1.1\%, primarily due
to the muon energy cut.  The statistical error on the total rate of
\nrn{\ism{12}C}{}{\ism{12}B} found in \sect{sec:b12analysis} applies, giving an
additional systematic of $^{+3.0}_{-1.2}$\% relative.

\item For results stated as a transition rate, the entire 1.5\% efficiency
error from \sect{sec:gammaselection} applies, as does the error on the number
of atomic captures on \is{12}{C}, 1.4\% (see \sect{sec:ratesloose}), plus a
very small 0.1\% error from the \mum lifetime in \is{12}{C}.

\item An additional overall systematic on each level is assigned for
uncertainty in the energy reconstruction following a muon.  The relative
uncertainties in the event counts from this effect are 0.5\% for 953\,keV,
2.8\% for 1674\,keV, 0.2\% for 2621\,keV and 1.7\% for 3597\,keV. 

\item To account for uncertainties in the number of \hn captures assumed in
\sect{sec:ncontam}, a further systematic error is assigned to the two adjacent
levels.  This is evaluated to be 1\% relative for 2621\,keV and 3\% relative
for 1674\,keV. 

\lu

All systematics combined are much smaller than the statistical error, which for
the most precise level is 12\% relative.

These results are compared to previous experimental and theoretical work in
\fig{fig:b12gcomp}.  Note particularly the comparison of these results to the
PQRPA model, which is also used to generate neutrino cross sections.  For
instance, for their \emph{parameterization III} (the bottom points shown for
Krmpoti{\'c} 2005) the ground state prediction is 4$\sigma$ high of our
measurement (experimental errors only), the 953\,keV level is 2$\sigma$ low,
the 1674\,keV level is 2$\sigma$ high, and the 2621 level in good agreement.

\section{Summary}\label{sec:conclusion}

Using the Double Chooz neutrino detector, the probabilities of final states of
muon capture on carbon, nitrogen and oxygen have been measured.  One set of
reactions, \nrn{\ism{12}C}{}{\ism{12}B^{(*)}}, has been observed before; more
precise results are reported here.  Other reactions have been observed for the
first time, and limits on several more have been set.  

The near detector is now also taking data.  Because it has an overburden of 120
meters water equivalent, it receives several times the flux of muons as
compared to the far detector.  This flux is also softer, giving a higher
fraction of stopping muons.  This provides an excellent data set for future
analysis of relatively short-lived isotopes. 

The RENO~\cite{reno} and Daya Bay~\cite{dayabay} detectors have similar
capabilities as Double Chooz and could also perform these measurements.
Similarly, a possibility to increase world data on muon capture in oxygen comes
from Super-Kamiokande, which has collected a large sample of stopping muons in
water~\cite{skspallation}, but has thus far treated them only as a background
to spallation studies.

The observed rate of \betan decays shows that stopping muons do not form a
significant background to $\theta_{13}$ measurements at any of the current
facilities.  Since the ratio of stopping to through-going muons rises as the 
amount of overburden decreases, it is possibly a relevant background for
inverse beta decay detectors with very small overburdens, such as those
investigating the reactor neutrino anomaly~\cite{reactoranomaly}.

\section{Acknowledgments} 
We thank the French electricity company EDF; the European fund FEDER;
the R\'egion de Champagne Ardenne; the D\'epartement des Ardennes;
and the Communaut\'e de Communes Ardenne Rives de Meuse.
We acknowledge the support of the CEA, CNRS/IN2P3, the computer centre CCIN2P3, and LabEx UnivEarthS in France (ANR-11-IDEX-0005-02);
the Ministry of Education, Culture, Sports, Science and Technology of Japan (MEXT) and the Japan Society for the Promotion of Science (JSPS);
the Department of Energy and the National Science Foundation of the United States;
U.S. Department of Energy Award DE-NA0000979 through the Nuclear Science and Security Consortium;
the Ministerio de Econom\'ia y Competitividad (MINECO) of Spain;
the Max Planck Gesellschaft, and the Deutsche Forschungsgemeinschaft DFG, the Transregional Collaborative Research Center TR27, the excellence cluster ``Origin and Structure of the Universe'', and the Maier-Leibnitz-Laboratorium Garching in Germany;
the Russian Academy of Science, the Kurchatov Institute and RFBR (the Russian Foundation for Basic Research);
the Brazilian Ministry of Science, Technology and Innovation (MCTI), the Financiadora de Estudos e Projetos (FINEP), the Conselho Nacional de Desenvolvimento Cient\'ifico e Tecnol\'ogico (CNPq), the S\~ao Paulo Research Foundation (FAPESP), and the Brazilian Network for High Energy Physics (RENAFAE) in Brazil.

\nolinenumbers

\bibliography{sisopaper}

\end{document}